\documentclass[twocolumn]{aastex631}

\pdfoutput=1 \usepackage{amsmath}
\usepackage[T1]{fontenc}
\usepackage{url}
\usepackage{graphicx}
\usepackage[caption=false]{subfig}
\usepackage{booktabs}

\begin{document}

\title{Investigating Optical Variability of the Blazar S5 0716+714 On Diverse Time-scales}

\author[0000-0001-5778-5679]{Erg\"un Ege}
\affiliation{Istanbul University, Institute of Graduate Studies in Science, 34116, Beyazit, Istanbul, T\"urkiye}

\author[0000-0003-1399-5804]{Aykut \"Ozd\"onmez}
\affiliation{Ataturk University, Faculty of Science, Department of Astronomy and Space Science, 25240, Yakutiye, Erzurum, T\"urkiye}

\author[0000-0003-4682-5166]{Aditi Agarwal}
\affiliation{Center for Cosmology and Science Popularization (CCSP) SGT University, Budhera, Delhi- NCR - 122006, India}

\author[0000-0002-0688-1983]{Tansel Ak}
\affiliation{Istanbul University, Faculty of Science, Department of Astronomy and Space Sciences, 34116, Beyazıt, Istanbul, T\"urkiye}

\begin{abstract}

We present the results of the observational study of the blazar S5 0716+716 in the optical bands $B$, $V$, $R$, and $I$ between March 2019 and August 2023 to investigate its variability on diverse time-scales. {The blazar was followed up by the T60 robotic telescope in Turkey for 416 nights to obtain long-term variability during this period. In order to search for intraday variability of the object, we have carried out 21 nights of observations with the T100 telescope for at least 1 hour.} The blazar showed a $\sim$ 2.47 mag variation in the optical R-band during our monitoring period, the brightest state on 18.01.2020 (MJD 58866) as R=12.109$\pm$0.011 and the faintest state on 23.03.2019 (MJD 58565) as R=14.580$\pm$0.013. We employed the nested ANOVA test and the power enhanced F-test to quantify intraday variability which showed that the blazar was significantly variable in the R-band on 12 out of 21 nights. Correlation analysis of the light curves shows that the emission in the BVRI optical bands was strongly correlated both in the short and long term without any time lag. The blazar has likely quasi-periods of 186$\pm30$, 532$\pm76$ days in the optical R-band light curve according to the WWZ, and the LS periodogram. The IDV and LTV features are discussed within the frame of prospective scenarios.
\end{abstract}


\section{Introduction} \label{sec:intro}

The Blazar class of Active Galactic Nuclei is comprised of BL Lacertae (BL Lac) objects and Flat Spectrum Radio Quasars (FSRQ). These objects are characterised by having rapid and high flux variability across the entire electromagnetic spectrum, from radio to gamma rays, as well as strong optical polarisation. In particular the BL Lac subclass is notable for its weak or lack of line profile in the optical spectra \citep{1995PASP..107..803U, 1995ARA&A..33..163W}. Having relativistic jets pointing close to the observer's line of sight ($<$10$^{\circ}$) makes their non-thermal radiation dominant \citep{2004NewAR..48..367K}. Like the other AGNs, the broadband continuum of the  Spectral Energy Distribution (SED) of blazars has two humps. The low energy part, from radio to X-ray, is synchrotron emission produced by relativistic electrons in the jet, the high energy part, from X-ray to $\gamma$-ray, is inverse Compton emission explained by various models; leptonic, hadronic or lepto-hadronic mixture \citep{2017SSRv..207....5R}.

The variability of blazars is classified based on the time-scale of the flux variations. Intraday Variation (IDV) or micro-variation refers to the variations in the flux of up to a few tenths of a magnitude that occur over a time-scale of a few minutes to a day \citep{1995ARA&A..33..163W} . Short-Term Variation (STV) refers to variability that are greater than one magnitude and occuring over a period of several months. Long-Term Variation (LTV) refers to changes in the flux of several magnitudes over a time-scale of months to many years \citep{2001A&A...374..435M, 2008AJ....135.1384G}.

The time-scales of the variability are associated with different radiation processes. Several scenarios have been proposed to explain the observed flux variations in blazars: instabilities in the accretion disk \citep{1996ASPC..110...42W}, the precession of the jet \citep{2011A&A...526A..51K}, shocks moving through the jet \citep{1996A&AS..120C.537M}, and changes in the Doppler factor due to relativistic emission plasma following a spiral path in the jet \citep{2017Natur.552..374R}, etc. The color change that accompanies the flux variation can lead to an elaboration of the structure of the source \citep{2002A&A...390..407V, 2016MNRAS.455..680A}. The analysis of flux variations and spectral behavior provides insights into the dynamics and time-scales of electron cooling in the blazar jets as well as facilitates the testing of theoretical models \citep{2008AJ....136.2359G, 2003A&A...402..151R}. Finding a trace of any periodicity in the flux variation provides clues about the accretion disk and central region, the SMBH and the jet \citep{2020MNRAS.492.5524O, Chatterjee_2018, Schmidt_2012}.

S5 0716+714 ($\alpha$ = 07h 21m 53.44s, $\delta$ = +71$^{\circ}$ 20' 36.4'') is one of the brightest BL Lac objects in the optical bands \citep{Rani2013}, and located at a distance of z = 0.2304 $\pm$ 0.0013 \citep{2023A&A...680A..52P}. It was first discovered in a 5 GHz survey of extragalactic radio sources in 1979 \citep{1981AJ.....86..854K}. It was detected in the very high energy (VHE, >100 GeV) $\gamma$-ray band by the MAGIC telescope during an optical ﬂare in 2008 \citep{2009ApJ...704L.129A}. {BL Lac objects are classified as low synchrotron peaked blazars (LSP, $\nu_{peak} \leq 10^{14}$ Hz), intermediate synchrotron peaked blazars (ISP, $10^{14}$ Hz $\leq \nu_{peak} \leq 10^{15}$ Hz), and high synchrotron peaked blazars (HSP, $\nu_{peak} \geq 10^{15}$ Hz) according to the peak frequency of the synchrotron hump \citep{Abdo_2010}. The blazar has a luminosity of 10$^{46}$ erg s$^{-1}$ and a peak in the UV-Xray band at a frequency of 10$^{14.96}$ Hz \citep{2016ApJS..226...20F}, which places it in the ISP class of blazars.}

Its brightness and extreme variability make it a popular target for multi-wavelength observations and time-domain astrophysical studies \citep[e.g.,][]{1996AJ....111.2187W, 2006A&A...451..797O, 2012MNRAS.425.1357G, 2014ApJ...783...83L, 2015ApJ...809..130C, 2015ApJS..218...18D}. The blazar had a brightness of 11.6 mag (optical R-band) which is the brightest value ever recorded on 18 Jan 2015 during its flare state \citep{2016MNRAS.455..680A}. The blazar was historically observed in its faintest state on 6 Jan 2014 (R=14.85$\pm0.06$ mag ) by \cite{2018AJ....156...36K}. Several studies have suggested different QPOs in the optical bands. \cite{2018AJ....155...31H} detected a QPO of 50 minutes. \cite{2023Ap&SS.368...88H} detected a QPO of 2.90$\pm$0.14 yr using 32 years of optical R-band data and associated the QPO with a probable binary SMBH. \cite{2022ApJ...938....8C} detected a transient QPO of $31.3\pm1.8$ days in the $\gamma$-ray band light curve of the blazar S5 0716+714 using weighted wavelet Z-transform (WWZ) and Lomb–Scargle periodogram (LSP). Likely QPOs of 24.24±1.09 days, 24.12±0.76 days, and 24.82±0.73 days in the optical V-, R-, and I-bands, respectively, were identified by \cite{2017A&A...605A..43Y}. In the optical and radio data of S5 0716+714 \cite{2003A&A...402..151R} detected a periodicity of 3.0$\pm$0.3 yr and a 5.5-6 yr respectively. This result is very close to that of \cite{Li_2023}, which is 2.63$\pm$0.22 yr in $\gamma$-ray. Using the auto-correlation function (ACF) analysis, \cite{Liao_2014} identified variability time-scales of $\sim$60-90 days across 14.5 GHz radio, optical V , X-ray, and $\gamma$-ray light curves and $\gamma$-ray and optical bands are found to be correlated during the flares. \cite{2013A&A...558A..92B} obtained a 72-hour continuous optical light curve with WEBT in February 2009, which showed no periodicity.

The spectral behavior of the blazar S5 0716+714 has been studied by many researchers. Some studies have claimed that the source has the property of being bluer when it is brighter, both on short- and long-term time-scales \citep{1997A&A...327...61G, 2008A&A...481L..79V, 2011ApJ...731..118C, 2016MNRAS.455..680A, 2016ApJ...831...92B, 2017AJ....154...42H, 2022ApJ...928...86G} while others have found the opposite \citep{2006MNRAS.366.1337S, 2012ARep...56..275V} or monitored either \citep{2003A&A...402..151R}.

In this paper we present the comprehensive variability results of the optical observational study of the blazar S5 0716+714 on diverse time-scales. The paper is structured as follows: Section 2 provides brief information of the observations, the telescopes used, and the data reduction process. Section 3 describes the statistical analysis techniques used to quantify the variability, investigate the correlations, and detect the periodicity. The results obtained are presented in section 4. The results are discussed and conclusions are drawn in section 5.

\section{Observations and Data Reduction} \label{sec:obs}

The photometric observations of S5 0716+714 were carried out in the optical BVRI bands with two optical telescopes T60 and T100 at the campus of the National Observatory of T\"urkiye located in Bakirlitepe, Antalya, T\"urkiye. The T100 telescope has a 1 m mirror and it is equipped with a cryogenic cooled CCD (model SI 1100 Cryo, size 4096$\times$4037 px). The T60 is a robotic telescope with a 60 cm mirror and a thermally cooled CCD (model Andor iKon-L, size 2048$\times$2048 px).

The observation period for the T100 and T60 telescopes was between March 2019 and August 2023. The T100 telescope was used to collect data for IDV analysis, allocating at least one hour and more in each observation night, while the T60 robotic telescope followed the target by taking at least two frames in each optical band filter as many nights as possible, depending on the weather conditions. {Exposure times range from 2 to 120 seconds and 15 to 120 seconds with the T100 and T60 telescopes respectively, depending on the filter, and the brightness of the blazar.}

Calibration of the raw images taken by the T100 telescope was done using standard Python libraries (e.g. astropy, ccdproc) using bias and flat frames taken regularly by the observatory's technical stuff. Since the CCD detector on the T100 telescope is cooled up to $-95$C, no dark calibration was performed on the frames taken by the T100. The Bias, Dark, and Flat fielding calibrations of the frames taken by the T60 robotic telescope were performed by the observatory pipeline. To obtain the instrumental magnitudes of the S5 0716+714 and the standard comparison stars, aperture photometry was performed on the nightly frames within aperture radii between 1 $\times$ FWHM and 5 $\times$ FWHM  to obtain the best signal to noise ratio. The finding chart of the blazar is taken from the study of \citet{1998A&AS..130..305V} and the brightness of the standard stars are taken from \citet{1998A&AS..130..305V, 1997A&A...327...61G}. {We provide our observational data, including instrumental magnitudes, exposure times, and dates, in the online data. We have taken into account magnitudes with an error less than 0.1 mag resulting in a dataset with an average error of 0.051 mag.} We have used star 5 as a reference, and stars 3 and 6 as comparisons for the evaluation of the blazar's flux, because of their high signal-to-noise ratio, close position to the source, close magnitude, and comparable color characteristics, which mitigate possible discrepancies in the photon statistics of the photometric measurements. The flux values are reliable because the target and the standard stars are very close together in the same FOV, i.e. have the same air mass and measurement conditions. {After applying the galactic extinctions given in the Nasa Extragalactic Database (NED)\footnote{https://ned.ipac.caltech.edu} as $A_B=0.112$, $A_V=0.085$, $A_R=0.067$, $A_I=0.047$ \citep{2011ApJ...737..103S}, the resulting magnitudes were converted to the corresponding fluxes following \cite{1998A&A...333..231B}.}

\section{Intra-Day Variability}\label{sec:analyses}

\subsection{Flux Variability}
\begin{figure*}[ht!]
\includegraphics[width=\linewidth]{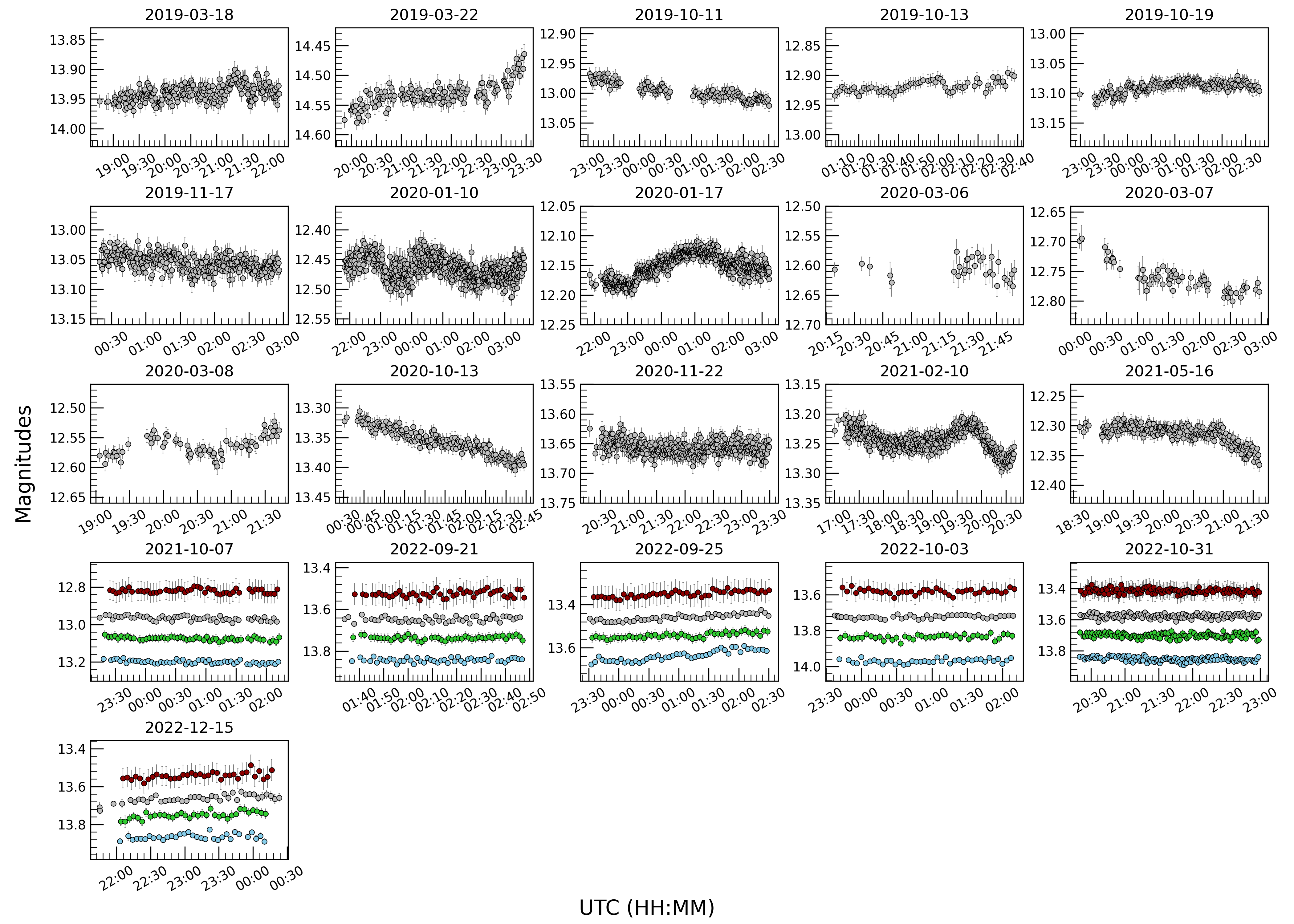}
\caption{Intraday light curves of the blazar S5 0716+714 in the optical $BVRI-$bands. For better visualisation, the dark red, silver, green and blue circles represent the B-,V-, R- and I-band magnitudes with an offset of -0.6, -0.28, 0.0, and +0.35, respectively. The dates of the observations are given at the top of each plot. \textbf{The complete observational log and results in the B, V, R, and I bands are available as the data behind the figure in the online Journal.}}
\label{fig:IDV_R}
\end{figure*}

To study the flux variability of blazars at different frequencies is essential to delineate both the size and the structure of the emission regions and the radiative process. Only those nights with more than one hour of observations were analysed to ensure that there were enough data points to detect meaningful IDV. We have 21 nightly observations that meet this criterion, of which 6 are multi-band (BVRI) and 15 are R-band only. The light curves of these 21 night observations are shown in the Figure \ref{fig:IDV_R}.

Various statistical methods have been used to quantitatively characterise variable sources, depending on the data to be analysed. To reveal the variability characteristics of the blazar in the optical BVRI bands on the intraday time-scale, we used the most common and robust statistical techniques; i.e. the power-enhanced F-test and the nested ANOVA test.

The power-enhanced F-test is a tool for comparing the blazar light curve with the combined variance of the comparison stars, which has been used in recent studies to detect IDV quantitatively \citep[e.g.][]{2015MNRAS.452.4263G, 2016MNRAS.460.3950P, 2019ApJ...871..192P, 2022ApJ...933...42A}. The power-enhanced F-statistic is defined as:
\begin{equation}
F_{enh}=\frac{s_{blz}^2}{s_c^2}
\end{equation}
where $s_\text{blz}^2$ is the  estimated variance difference between the blazar and the reference star, $s_\text{c}^2$ represents the combined variance of the comparison star and the reference star calculated with Equation (13) in \citet{2014AJ....148...93D}.

In our research, we have used the field star S5 as a reference star which is the reference star with the closest magnitude to the blazar. The field stars S3 and S6 are used as comparison stars (k = 2). Since the comparison stars and the blazar are in the same field, we have the same number of observations for each (N), the F-statistics have $\mu_1 = N-1$ and $\nu_2 = k(N-1)$ degrees of freedom in numerator and denominator, respectively. We have estimated the $F_\text{enh}$  and compared it with the critical value ($F_c$) at $\alpha = 0.01$, corresponding to a 99\% confidence level. A light curve is classified as variable (V) if $F_\text{enh}\geq F_c$, and otherwise as non-variable (NV).

The ANOVA (for Analysis of Variance) is a tool for evaluating the mean values of the dispersion between groups of observations. The Nested ANOVA is a modified version of ANOVA that allows the use of multiple reference stars to generate differential LCs of the blazar. Since the nested ANOVA does not require a comparison star, all reference stars are used in the test.
In our study we used three reference stars (S3, S5 and S6) to generate differential LCs of the blazar. The differential \phantom{two} LCs are divided into groups of five points each. The mean square due to the groups ($MS_G$) and the nested observations in the groups ($MS_{O_{G}}$) were calculated according to the Equation (4) from \citet{2015AJ....150...44D}  \citep[see also][Chapter 14]{montgomery2012design}.

The resulting ratio $F = MS_G/MS_{O_{G}}$ corresponds to an $F$-distribution with $(a-1)$ and $(a(b-1))$ degrees of freedom, in the numerator and denominator, respectively. For a significance level of $\alpha = 0.01$, the light curve is classified as variable (V) if the $F$-statistic is greater than or equal to the critical value ($F_c$), otherwise it is non-variable (NV).

To quantify the variability amplitude, we used the following equation;

\begin{equation}
A = 100 \times \sqrt {(A_{max} - A_{min})^{2} - 2\sigma^{2}} \ [\%]
\end{equation}

where $A_{max}$ and $A_{min}$ are the maximum and the minimum values of the measured magnitudes over the entire observation period, and $\sigma$ is the mean error \citep{1996A&A...305...42H}.

Table \ref{tab:vartest} shows the results of the $F_\text{enh}$-tests and nested ANOVA tests. A light curve is classified as variable (V) in the table only if significant variations were confirmed by both tests, otherwise it is classified as non-variable (NV). Variability amplitudes were also calculated for the nights, showing IDV.

\begin{table*}[ht!]
\caption{The test results of IDV of the blazar S5 0716+714.}
\label{tab:vartest}
\centering
\begin{tabular}{lccccccccccc}
\hline
\hline
Obs. date &  Band & Avg. Mag. & $t_\text{obs}$ & \multicolumn{3}{c}{Power-enhanced F-test}  &  \multicolumn{3}{c}{Nested ANOVA test}& Status & Amplitude\\
\cline{5-7}\cline{8-10}
 yyyy-mm-dd &   &  & hours &DOF($\nu_1$,$\nu_2$) &  $F_{enh}$ &  $F_c$ & DOF($\nu_1$,$\nu_2$) &  $F$ &  $F_c$  &  & \% \\
\hline
2019-03-18 &    R &	13.939 &      3.46 &     (239, 478) &    0.54 &        1.29 &        (47, 192) &     5.00 &        1.65 &     NV &        ... \\
2019-03-22 &    R &	14.531 &      3.60 &     (140, 140) &    2.45 &        1.48 &        (27, 112) &    12.65 &        1.91 &      V &      11.23 \\
2019-10-11 &    R &	12.997 &      3.48 &     (130, 130) &    5.39 &        1.51 &        (25, 104) &    52.39 &        1.96 &      V &       5.10 \\
2019-10-13 &    R &	12.919 &      1.50 &       (67, 67) &    3.02 &        1.78 &         (12, 52) &    12.00 &        2.55 &      V &       3.73 \\
2019-10-19 &    R &	13.090 &      3.80 &     (169, 338) &    0.08 &        1.35 &        (33, 136) &    32.24 &        1.81 &     NV &        ... \\
2019-11-17 &    R &	13.055 &      2.62 &     (319, 638) &    1.59 &        1.25 &        (63, 256) &     5.44 &        1.55 &      V &       7.01 \\
2020-01-10 &    R &	12.466 &      5.80 &    (606, 1212) &    1.28 &        1.18 &       (120, 484) &     7.78 &        1.38 &      V &       9.41 \\
2020-01-17 &    R &	12.151 &      5.35 &    (598, 1196) &    4.63 &        1.18 &       (118, 476) &    52.52 &        1.38 &      V &       8.40 \\
2020-03-06 &    R &	12.607 &      1.58 &       (30, 60) &    1.10 &        2.03 &          (5, 24) &     4.80 &        3.90 &     NV &        ... \\
2020-03-07 &    R &	12.762 &      2.90 &      (64, 128) &    3.84 &        1.63 &         (12, 52) &    26.54 &        2.55 &      V &      10.08 \\
2020-03-08 &    R &	12.564 &      2.66 &      (71, 142) &    3.12 &        1.59 &         (13, 56) &    19.67 &        2.47 &      V &       7.20 \\
2020-10-13 &    R &	13.356 &      2.21 &     (168, 336) &    5.92 &        1.36 &        (32, 132) &    72.12 &        1.82 &      V &       9.41 \\
2020-11-22 &    R &	13.657 &      3.18 &     (366, 732) &    0.77 &        1.23 &        (72, 292) &     2.54 &        1.51 &     NV &        ... \\
2021-02-10 &    R &	13.244 &      3.66 &     (430, 860) &    2.96 &        1.21 &        (85, 344) &    21.32 &        1.46 &      V &       9.19 \\
2021-05-16 &    R &	12.314 &      3.00 &     (209, 418) &    2.64 &        1.31 &        (41, 168) &    39.22 &        1.71 &      V &       7.46 \\
2021-10-07 &    B &	13.800 &      2.89 &      (53, 106) &    1.27 &        1.71 &          (9, 40) &     4.06 &        2.89 &     NV &        ... \\
2021-10-07 &    V &	13.355 &      2.89 &      (52, 104) &    0.95 &        1.72 &          (9, 40) &     6.72 &        2.89 &     NV &        ... \\
2021-10-07 &    R &	12.966 &      2.94 &      (53, 106) &    1.50 &        1.71 &          (9, 40) &     5.08 &        2.89 &     NV &        ... \\
2021-10-07 &    I &	12.470 &      2.77 &       (49, 98) &    0.74 &        1.74 &          (9, 40) &     4.80 &        2.89 &     NV &        ... \\
2022-09-21 &    B &	14.443 &      1.17 &       (47, 94) &    0.73 &        1.76 &          (8, 36) &     0.74 &        3.05 &     NV &        ... \\
2022-09-21 &    V &	14.016 &      1.16 &       (47, 94) &    0.67 &        1.76 &          (8, 36) &     2.92 &        3.05 &     NV &        ... \\
2022-09-21 &    R &	13.647 &      1.20 &       (49, 98) &    0.80 &        1.74 &          (9, 40) &     1.89 &        2.89 &     NV &        ... \\
2022-09-21 &    I &	13.176 &      1.16 &       (48, 96) &    0.96 &        1.75 &          (8, 36) &     0.70 &        3.05 &     NV &        ... \\
2022-09-25 &    B &	14.237 &      2.92 &       (47, 94) &    3.94 &        1.76 &          (8, 36) &    43.68 &        3.05 &      V &       7.88 \\
2022-09-25 &    V &	13.822 &      2.92 &       (47, 94) &    1.94 &        1.76 &          (8, 36) &    17.69 &        3.05 &      V &       4.41 \\
2022-09-25 &    R &	13.459 &      2.98 &       (48, 96) &    2.43 &        1.75 &          (8, 36) &    22.10 &        3.05 &      V &       5.57 \\
2022-09-25 &    I &	13.000 &      2.92 &       (47, 94) &    2.32 &        1.76 &          (8, 36) &    11.13 &        3.05 &      V &       5.45 \\
2022-10-03 &    B &	14.569 &      2.43 &       (39, 78) &    1.15 &        1.86 &          (7, 32) &     1.22 &        3.26 &     NV &        ... \\
2022-10-03 &    V &	14.115 &      2.43 &       (40, 80) &    0.91 &        1.85 &          (7, 32) &     1.54 &        3.26 &     NV &        ... \\
2022-10-03 &    R &	13.723 &      2.53 &       (43, 86) &    0.46 &        1.81 &          (7, 32) &     1.67 &        3.26 &     NV &        ... \\
2022-10-03 &    I &	13.231 &      2.43 &       (40, 80) &    1.23 &        1.85 &          (7, 32) &     1.34 &        3.26 &     NV &        ... \\
2022-10-31 &    B &	14.454 &      2.64 &     (118, 236) &    1.13 &        1.44 &         (22, 92) &     3.96 &        2.04 &     NV &        ... \\
2022-10-31 &    V &	13.981 &      2.64 &     (119, 238) &    0.82 &        1.43 &         (23, 96) &     1.00 &        2.01 &     NV &        ... \\
2022-10-31 &    R &	13.576 &      2.64 &     (122, 244) &    0.74 &        1.43 &         (23, 96) &     1.73 &        2.01 &     NV &        ... \\
2022-10-31 &    I &	13.065 &      2.64 &     (116, 232) &    0.71 &        1.44 &         (22, 92) &     1.37 &        2.04 &     NV &        ... \\
2022-12-15 &    B &	14.465 &      2.12 &       (32, 64) &    0.82 &        1.98 &          (5, 24) &     3.82 &        3.90 &     NV &        ... \\
2022-12-15 &    V &	14.030 &      2.12 &       (34, 68) &    1.76 &        1.95 &          (6, 28) &     4.90 &        3.53 &     NV &        ... \\
2022-12-15 &    R &	13.664 &      2.63 &       (39, 78) &    1.56 &        1.86 &          (7, 32) &     7.06 &        3.26 &     NV &        ... \\
2022-12-15 &    I &	13.193 &      2.18 &       (35, 70) &    1.09 &        1.93 &          (6, 28) &     3.02 &        3.53 &     NV &        ... \\
\hline
\end{tabular}
\end{table*}

{The source was variable on minute time-scales on seven nights with a variability amplitude between 3.73 and 11.23.} Although multi-band LCs are available for the 2022-09-25 and 2022-12-15, only one band (B-band and V-band respectively) of the light curve showed variability. Duty cycle (DC) is commonly used as a quantitative measure to assess the likelihood of variability of a source. The DC is defined as the fraction of nights showing variability out of the total number of nights \citep{1999A&AS..135..477R}.
\begin{equation}
    DC=100 \frac{\sum_{i=1}^n (N_i /\Delta t_i)}{\sum_{i=1}^n (1 /\Delta t_i)} \%
\end{equation}
where $\Delta t_i = \Delta t_{i,\text{obs}} (1+z)^{-1}$ is the duration of monitoring from the reference frame of the source. $N_i$ takes a value of 1 if the source is marked as intraday variable, otherwise it takes a value of 0. Out of 21 observation nights between March 2019 and August 2023, the blazar was monitored for at least 1, 2, and 3 hours on 21, 18, and 9 nights, respectively.

\begin{table}[ht!]
\caption{Duty cycle of the intraday observations within different observation duration. $N_{V,T}$ is the number of variable and total nights, respectively.}
\label{tab:DC}
\centering
\begin{tabular}{ccc|cc|cc}
\hline \hline 
 Band &  \multicolumn{2}{c|}{$t_\text{obs}\geq1$ hr}   & \multicolumn{2}{c|}{$t_\text{obs}\geq2$ hr} & \multicolumn{2}{c}{$t_\text{obs}\geq3$ hr} \\
 \cline{2-3}\cline{4-5}\cline{6-7}
  &  $N_{V,T}$ & DC &  $N_{V,T}$ & DC\ &  $N_{V,T}$ & DC \\
\hline 
 B & 1, 6 & 12\% &  1, 5 & 18\% & - & - \\
 V & 1, 6 & 12\% &  1, 5 & 18\% & - & - \\
 R & 12, 21 & 52\% & 11, 18 & 59\% & 6, 9 & 64\% \\
 I & 1, 6 & 12\% & 1, 5 & 18\% & - & - \\
 \hline 
\multicolumn{7}{c}{For IDV with brighter than average LTV mag.}\\
\hline 
R & 8, 10 & 74\% & 7, 8 & 86\% & 4, 4 & 100\% \\
\hline 
\end{tabular}
\end{table}

The duty cycle of IDV of the blazar S5 0716+714 (DC) has been determined in a wide range from 30\% up to 100 \% in previous studies \citep{2011ApJ...731..118C, 2016MNRAS.455..680A, 2018AJ....156...36K, 2023MNRAS.tmp.3427T}. Based on our observations, we obtained a value of up to 64\% of DC in different bands and different observation durations. \cite{2018AJ....155...31H} monitored the source for less than 1 h and reported a DC of 19.57\%, and, in another study carried out over 13 nights in January–February 2012, a DC of 44\% was estimated when the source was monitored for about 5 hr \citep{2017AJ....154...42H}. In order to check for any correlation between the IDV of the source and the monitoring time allocated, we calculated the DC for 1, 2 or 3 hours of monitoring period (Table \ref{tab:DC}). The results confirm the well-known fact that the longer the monitoring time, the higher the probability of detecting variability, i.e., a higher DC \citep{2005A&A...440..855G, 2010MNRAS.404.1992R, 2018AJ....155...31H}. We have also calculated the DC values for the nights with magnitudes brighter than the average LTV magnitudes. The calculated DC values are dramatically increased for these nights (Table \ref{tab:DC}), suggesting intraday variability related to the state of the blazar.

\subsection{Minimum Time-scale}
\begin{figure*}[ht!]
\includegraphics[width=\linewidth]{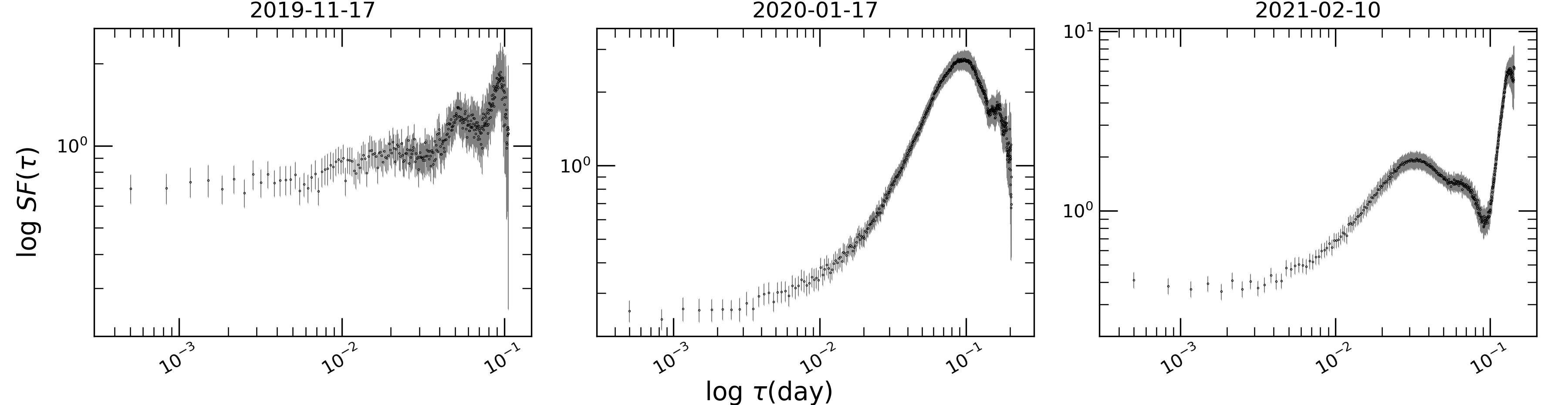}
\caption{Structure Function of the selected Intraday light curves of the blazar S5 0716+714 in the optical $R-$band. The dates of observation are written at the top of each plot.}
\label{fig:IDV_sf}
\end{figure*}

The Structure Function (SF) is a method of indicating how much a source is variable in a given time lag \citep{1985ApJ...296...46S} and is commonly used to quantify the time-scale of variation of LCs where the data points are not evenly sampled datasets. It is a function that gives the square of the mean difference between the flux densities corresponding to the time-scale. We use first-order SF, which is defined as:
\begin{equation}
    SF(\tau_i)=\frac{1}{N}\sum_{}^{}[M(t+\tau_i)-M(t)]^2
\end{equation}
where M(t) is the magnitude at a given time t and $\tau$ is the time lag.
Since the blazars show variability on all time-scales, a typical SF graph will show a linear increase and end with a plateau (or saturation) unless there is no periodicity, flare or flicker. Local minima following local maxima may indicate a repeating pattern, and the time between dips gives the periodicity. If SF has no plateau at the end, the maximum time-scale of variability is longer than the monitoring time. On the other hand a local minimum following a dip gives minimum time-scale as seen in Figure \ref{fig:IDV_sf}, and this finding can be used to place a minimum limit on the radiative size and mass of the SMBH.
We obtained the SF of the nightly observations using the code described in \cite{2018MNRAS.478.2557G} and the SF plots of the days with the most significant variations.{ On 10 Feb 2021, SF reaches to a local maximum at $0.0333 \pm 0.0035$ d ($\sim$ 48 min) and then has a dip after another rapid rise.} On 17 Jan 2021 SF reaches to $0.0982 \pm 0.0018$ ($\sim$ 141 min) and then declines. On 17 Nov 2019 there is more than one local maximum and dip which can be interpreted as flickering.

Using the minimum time-scale of an accreting black hole \citep{1974ApJ...192L...3E} we can derive the limit of the emission region by the equation
\begin{equation}
R \geq \frac{\delta c \Delta t_{var}}{1+z} 
\end{equation}
where $\delta$ is the Doppler factor, c is the speed of light, $\Delta$ t is the minimum time-scale, z is the redshift of the blazar which is 0.2304. Considering that the Doppler factor $\delta$ has been taken to be between 4 and 15 in the previous studies. Assuming $\delta$ to be 15, the maximum size of the emission region can be estimated to be of the order of $\approx 10^{15} cm$. The minimum time-scale of the variability has been used to estimate the mass of the central SMBH of the blazar in numerous studies \citep{2008AJ....136.2359G, 2010MNRAS.404.1992R, 2011ApJ...731..118C, 2015ApJS..218...18D, 2016MNRAS.455..680A, 2018AJ....156...36K} by using the following equation in the study of \cite {1982Natur.300..506A} as below:

\begin{equation}
    M = 1.62 \times 10^4 \frac{\delta \Delta t}{1+z} M_\circ
\end{equation}

Assuming $\delta$ is 15, the mass of the SMBH is estimated to be $5.69 \times 10^8 M_\circ$ (assuming the SMBH is a Kerr black hole).

\subsection{Spectral Variability}\label{sec:spec_var}

\begin{figure*}
    \centering
    \includegraphics[width=0.99\linewidth]{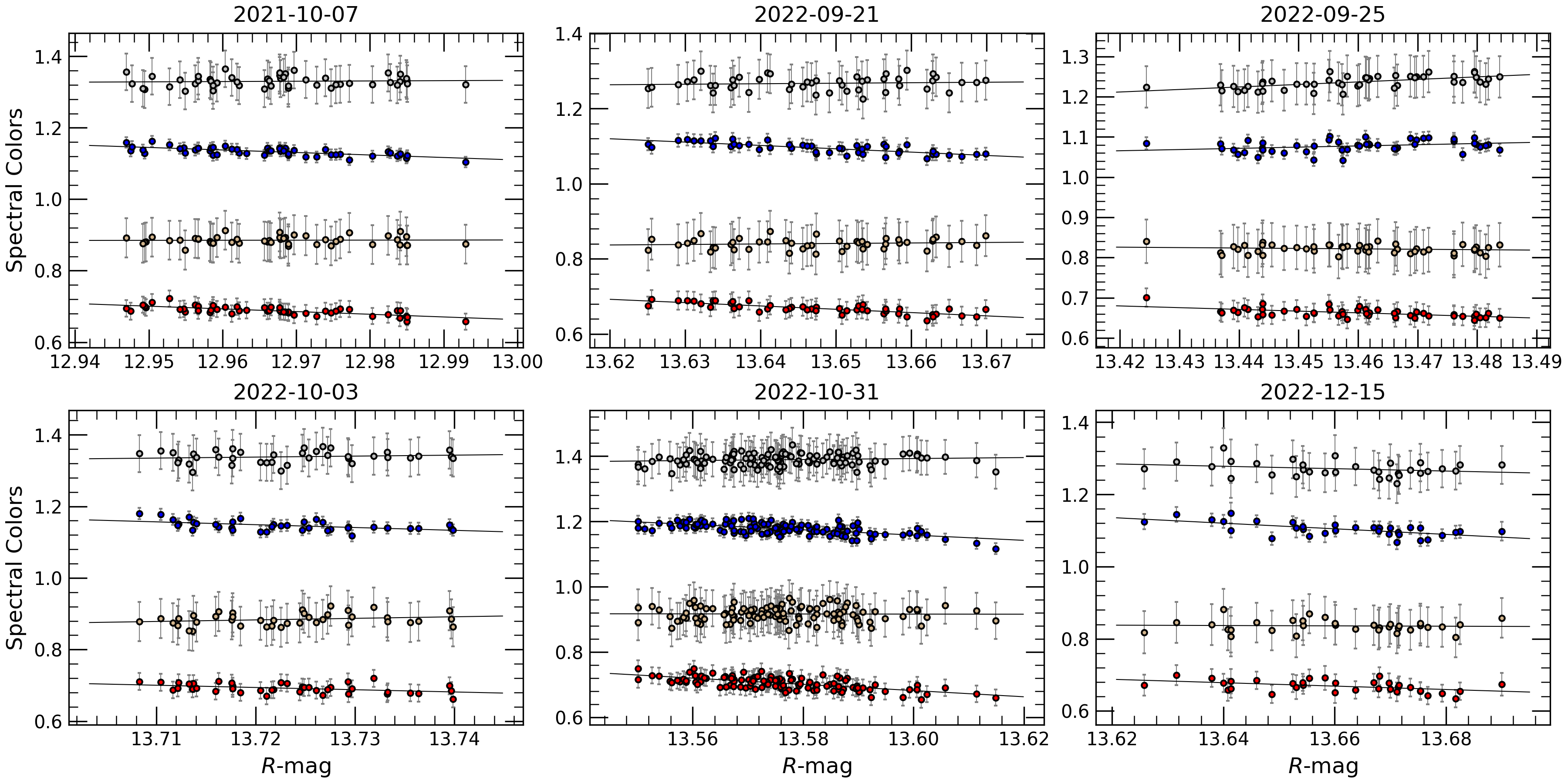}
    \caption{The intraday color–magnitude plots of the blazar S5 0716+714. The blue, silver, dark red and light brown colored dots represent $B-R$, $B-I$, $V-R$ and $V-I$ color indices. We have added +0.3 offset values to the $B-R$ and $V-R$ colors in order to distinguish between the distributions. Black lines represent linear fits {and their} parameters are given in the Table \ref{tab:IDV_cmd_fit}. }
    \label{fig:IDV_cmd}
\end{figure*}

Optical flux variations of blazars are often associated with spectral changes. {To investigate the spectral variability of the blazar S5 0716+714 on intra-night time-scales, we have plotted the spectral colors against the R-band magnitude for each night.} An example is shown in Figure \ref{fig:IDV_cmd}. For the configuration of the spectral colors, we have chosen the color indices $B-R$, $B-I$, $V-R$ and $V-I$ with the larger frequency base. In order to quantitatively determine the correlation between the spectral colors and brightness, we performed a regression analysis by fitting a straight line, using the linear model $y = mx + c$ (y = color index, x = mean magnitude), and extracted the parameters slope (m), intercept (c), correlation coefficient (r), p-value given in Table \ref{tab:IDV_cmd_fit}. We couldn't find a strong color relation that the blazar shows an achromatic behavior on the intraday scale. We should note that our multi-band night observations were made when the blazar was in a relatively quiescent state. 

\begin{table}[ht!]
\centering
\caption{Linear fits to Colors vs R-band magnitude for IDV data.}
\label{tab:IDV_cmd_fit}
\scriptsize
\begin{tabular}{lccccc}
\toprule
Date &   Color &             m &             c &      r &         p \\
\midrule
2021-10-07 &   $B-R$ & $-0.70\pm0.10$ & $ +9.94\pm1.30$ & $-0.70$ & 6.53e-09 \\
 &   $B-I$ & $+0.08\pm0.19$ & $ +0.23\pm2.45$ & $+0.07$ & 6.57e-01 \\
 &   $V-R$ & $-0.75\pm0.10$ & $+10.15\pm1.34$ & $-0.72$ & 2.41e-09 \\
 &   $V-I$ & $+0.02\pm0.15$ & $ +0.60\pm1.91$ & $+0.02$ & 8.82e-01 \\
2022-09-21 &   $B-R$ & $-0.89\pm0.12$ & $+12.93\pm1.62$ & $-0.74$ & 1.95e-09 \\
 &   $B-I$ & $+0.13\pm0.22$ & $ -0.56\pm2.95$ & $+0.09$ & 5.39e-01 \\
 &   $V-R$ & $-0.88\pm0.10$ & $+12.43\pm1.41$ & $-0.79$ & 5.50e-11 \\
 &   $V-I$ & $+0.13\pm0.17$ & $ -0.98\pm2.29$ & $+0.12$ & 4.30e-01 \\
2022-09-25 &   $B-R$ & $+0.30\pm0.14$ & $ -3.26\pm1.86$ & $+0.30$ & 3.56e-02 \\
 &   $B-I$ & $+0.63\pm0.12$ & $ -7.19\pm1.59$ & $+0.61$ & 3.34e-06 \\
 &   $V-R$ & $-0.43\pm0.09$ & $ +6.15\pm1.17$ & $-0.59$ & 1.01e-05 \\
 &   $V-I$ & $-0.10\pm0.10$ & $ +2.22\pm1.34$ & $-0.15$ & 3.01e-01 \\
2022-10-03 &   $B-R$ & $-0.79\pm0.22$ & $+11.69\pm2.95$ & $-0.52$ & 7.60e-04 \\
 &   $B-I$ & $+0.27\pm0.31$ & $ -2.42\pm4.31$ & $+0.14$ & 3.90e-01 \\
 &   $V-R$ & $-0.63\pm0.22$ & $ +8.99\pm3.04$ & $-0.42$ & 7.35e-03 \\
 &   $V-I$ & $+0.44\pm0.31$ & $ -5.11\pm4.25$ & $+0.22$ & 1.67e-01 \\
2022-10-31 &   $B-R$ & $-0.81\pm0.10$ & $+11.82\pm1.30$ & $-0.61$ & 1.15e-13 \\
 &   $B-I$ & $+0.15\pm0.12$ & $ -0.66\pm1.68$ & $+0.12$ & 2.24e-01 \\
 &   $V-R$ & $-0.95\pm0.10$ & $+13.33\pm1.37$ & $-0.66$ & 4.12e-16 \\
 &   $V-I$ & $-0.02\pm0.15$ & $ +1.13\pm2.07$ & $-0.01$ & 9.18e-01 \\
2022-12-15 &   $B-R$ & $-0.78\pm0.17$ & $+11.41\pm2.28$ & $-0.64$ & 5.75e-05 \\
 &   $B-I$ & $-0.33\pm0.22$ & $ +5.81\pm3.04$ & $-0.26$ & 1.46e-01 \\
 &   $V-R$ & $-0.47\pm0.16$ & $ +6.80\pm2.15$ & $-0.47$ & 5.39e-03 \\
 &   $V-I$ & $-0.05\pm0.18$ & $ +1.49\pm2.52$ & $-0.05$ & 7.97e-01 \\
\bottomrule
\end{tabular}
\end{table}

\section{Long-Term Variablity (LTV)}

\subsection{Flux Variability}

\begin{figure*}
\fig{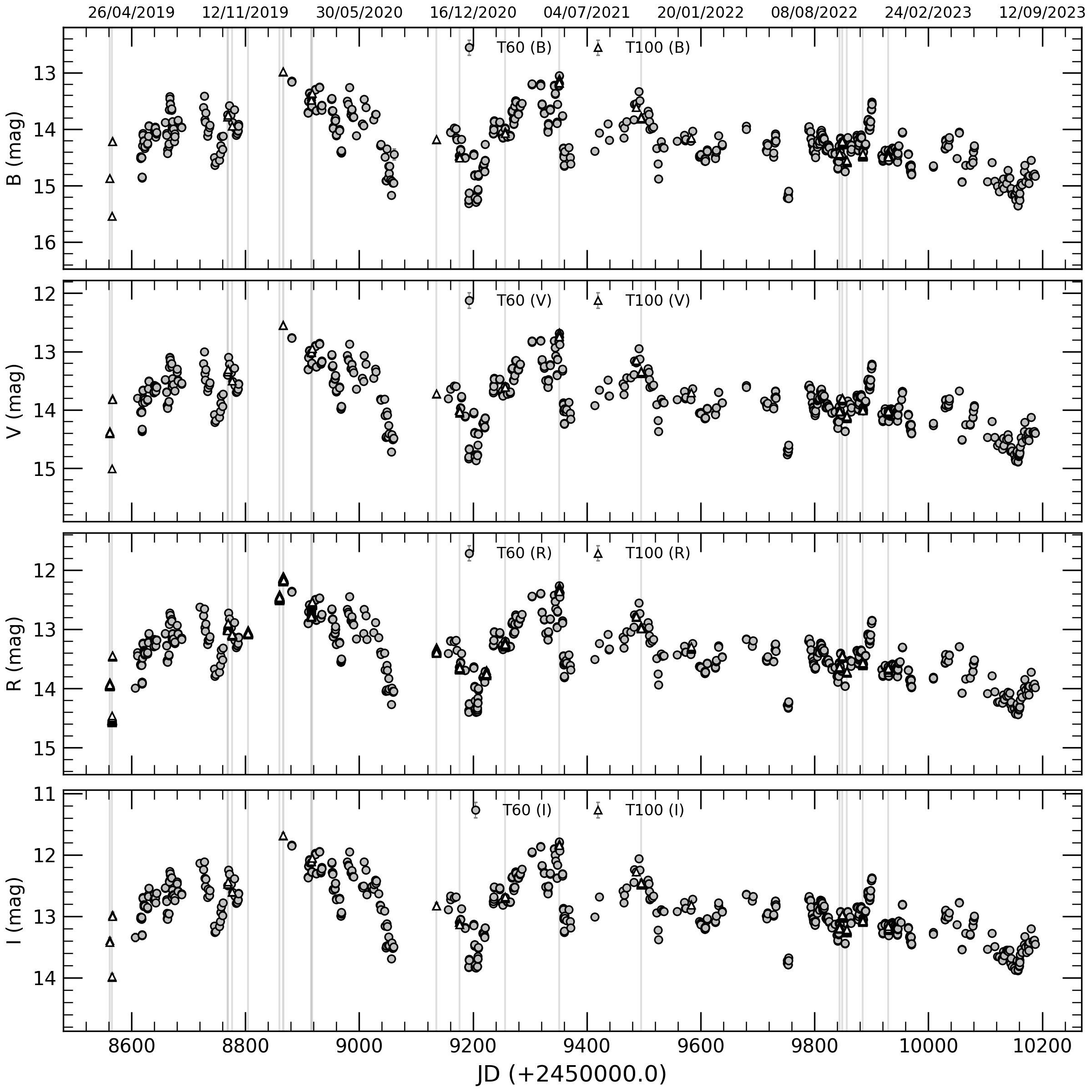}{0.95\textwidth}{}
\caption{Long-term light curves of the blazar S5 0716+714 in the optical $B$,$V$, $R$, and $I$ bands. Vertical lines represent the date of intraday observations as given in Fig \ref{fig:LTV_alpha_R}.}
\label{fig:LTV_BVRI}
\end{figure*}

The blazar S5 0716+714 was simultaneously followed-up in the optical BVRI bands for 416 nights between March 2019 and August 2023. The resulting long-term light curve can be seen in Figure \ref{fig:LTV_BVRI}. The brightness of the blazar in the optical R-band has varied between $12.109\pm0.011$ (on MJD 58866) and $14.580\pm0.013$ (on MJD 58565). The mean magnitude was $13.080\pm0.012$ as the variability amplitude was 247.043 \%. The brightness in the BVRI bands is summarised in the Table \ref{tab:LTV_result}. Considering the known periodicity and the historical brightness, we observed the object in both high and quiescent states. The blazar showed an NIR brightening in December 2019 \citep{2019ATel13359....1C} before we detected the highest state of our monitoring period on 18 January 2020. From early 2022 onwards, the source is likely to be in its quiescent phase, fainter than its average magnitude.

Blazars show long-term variability across the electromagnetic spectrum from radio to gamma. The optical flux variability can be explained by several intrinsic scenarios and should be studied together with the color and spectral behavior, and the periodicity. The precession of the jet can cause long-term variations in the observed flux \citep{2011A&A...526A..51K}. Since the emission mechanism remains the same, no change in the color behavior is expected and a (quasi)periodicity can coexist. The inhomogeneity of the jet, instability of the accretion rate are other mechanisms of LTV.

\begin{table*}
\caption{Results of the LTV analysis of the blazar S5 0716$+$714.}
\label{tab:LTV_result}                   
\centering 
\begin{tabular}{cccccc}       
\hline\hline                		 
Band &  Brightest magnitude/MJD & Faintest magnitude/MJD  & Average magnitude & Variability amplitude (\%)\\
\hline
B & $12.979\pm0.010$ / 58866.42621 & $15.540\pm0.016$ /	58565.33069 & $14.234\pm0.015$ & 256.123 \\
V & $12.542\pm0.020$ / 58866.42378 & $15.013\pm0.023$ /	58565.32583 & $13.821\pm0.022$ & 247.084 \\
R & $12.109\pm0.011$ / 58866.54426 & $14.580\pm0.013$ /	58565.33793 & $13.080\pm0.012$ & 247.043 \\
I & $11.687\pm0.050$ / 58866.42851 & $13.990\pm0.051$ /	58565.32040 & $12.910\pm0.051$ & 230.217 \\
\hline                          
\end{tabular}
\end{table*}

\subsection{Spectral Variability}

We employed a simple power law in the form of $\log(F_\nu)=-\alpha\log(\nu)+C$ to obtain the nightly spectral indices (F is the flux density, $\nu$ is the frequency, and $\alpha$ is the spectral index). The results of the fits are shown in the Table \ref{tab:sed}. The spectral index varies between 0.828 and 1.432 during our observation period and the mean spectral index is 1.092. These results are similar to the study of \cite{2021MNRAS.501.1100R} in where it varies from 0.95 to 1.34 (mean value of 1.14).

\begin{figure}
\includegraphics[width=\columnwidth]{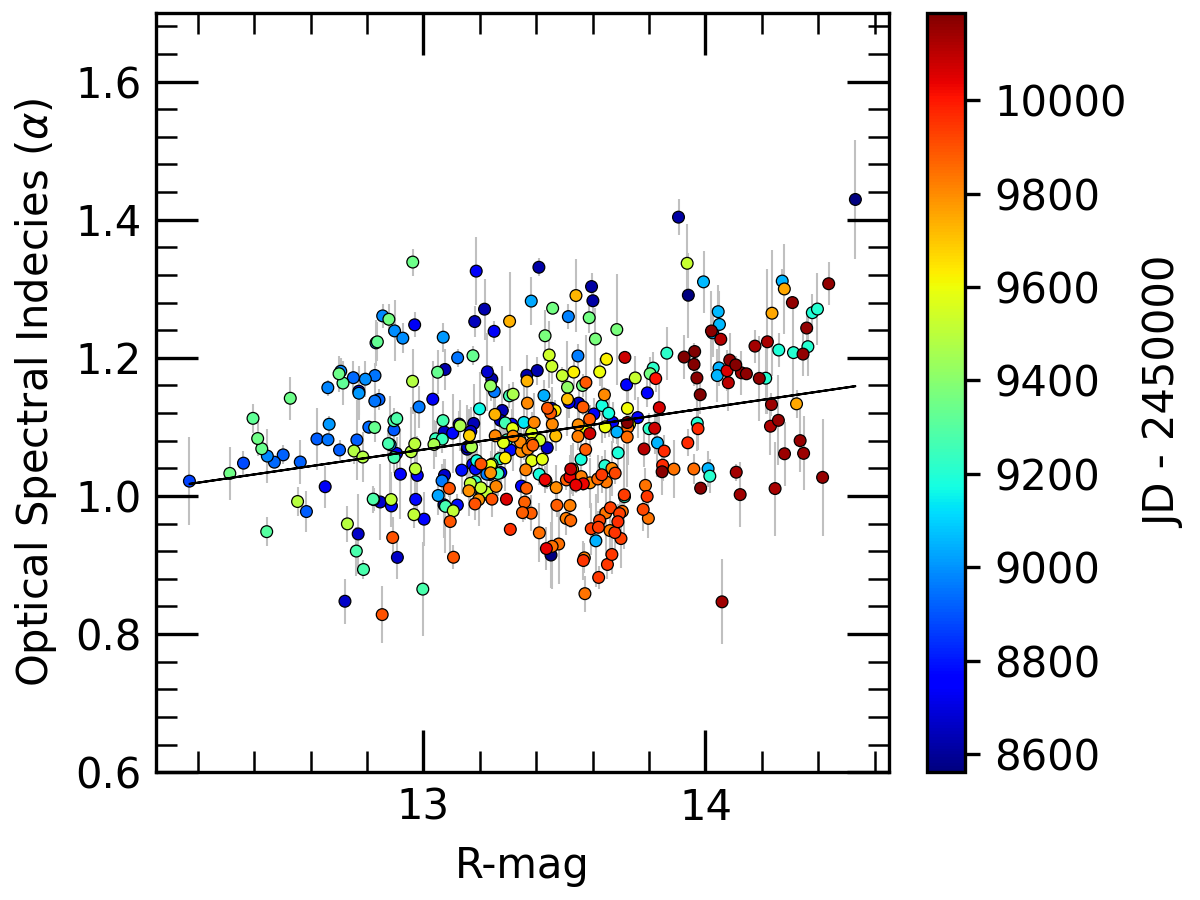}
\caption{Optical Spectral Indices vs R-mag plot of Long-Term Observations in Optical BVRI Bands}
\label{fig:LTV_alpha_R}
\end{figure}

To investigate the spectral behavior of the blazar S5 0716+714, we studied the variation of the spectral index with time and R-band magnitude. {We fitted a straight line to the plots of the spectral index (Fig. \ref{fig:LTV_alpha_R}) resulting in a slope of 0.06 with r=0.272 and p=1.93e-07. There is an aggregation of alpha between 13-14 mag when the brightness is in the medium state and the BWB behavior is also clearly visible in the plot.}

\begin{figure*}
\centering
\includegraphics[width=0.75\textwidth]{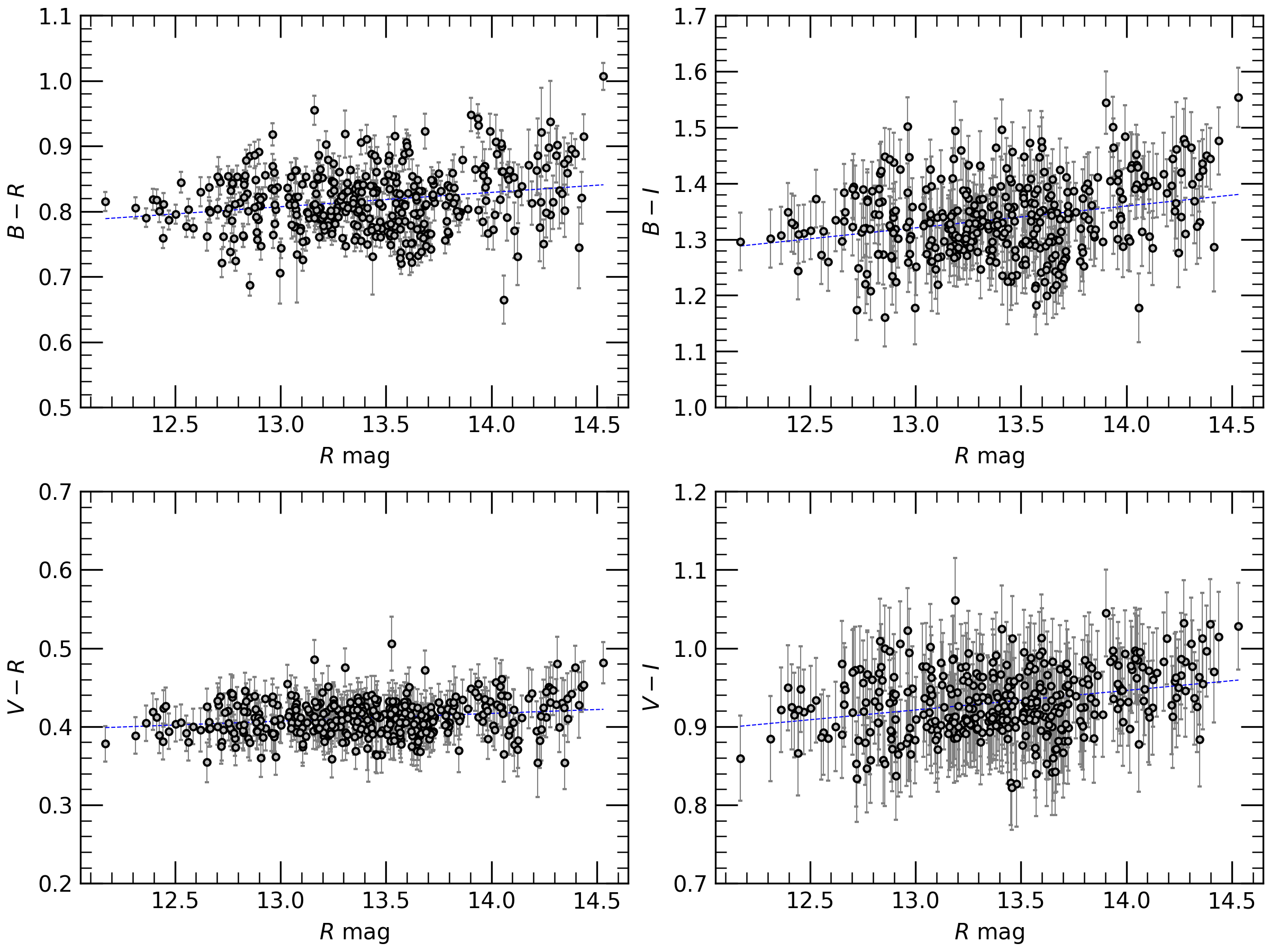}
\caption{Color indices of Long-Term LCs}
\label{fig:LTV_cmd}
\end{figure*}

We performed linear regression analysis on the spectral colors vs R-mag of the long-term LC (Fig. \ref{fig:LTV_cmd}) to reveal the color behavior of the source. {The slope values and the constants derived from these fits are listed in Table \ref{tab:LTV_fit}. Both in the Figure \ref{fig:LTV_alpha_R} and the Figure \ref{fig:LTV_cmd} show that the object has a BWB behavior as can be seen in the Table \ref{tab:LTV_fit}.} The value of the B-I color index varies between 1.161 and 1.553 (mean value 1.337) during the monitoring period. The variation in this value may be due to changes in the contribution of different emission components \citep{2016MNRAS.455..680A}. As the object becomes brighter, the jet emission dominates more the emission from the central components and the host galaxy, so it becomes bluer.

\begin{table}
\caption{Straight line fits to optical SEDs of the blazar S5 0716+714. $\alpha$ = spectral index and $C$ = intercept of log($F_{\nu}$) against log($\nu$); $r$ = Correlation coefficient; $p$ = null hypothesis probability} 
\label{tab:sed} 
\centering 
\scriptsize
\begin{tabular}{lcccc}
\hline\hline 
Date & $\alpha$ & $C$ & $r$ & $p$ \\
\hline
2019-03-18 &  $+1.291\pm0.061$ & $ -6.141\pm0.903$ & +0.998 &  2.256e-03 \\
2019-03-22 &  $+1.429\pm0.086$ & $ -4.348\pm1.269$ & +0.997 &  3.628e-03 \\
2019-03-23 &  $+0.914\pm0.048$ & $-11.459\pm0.699$ & +0.997 &  2.690e-03 \\
... &... &... &... &...  \\
... &... &... &... &...  \\
2023-08-27 &  $+1.209\pm0.006$ & $ -7.334\pm0.087$ & +1.000 &  2.412e-05 \\
2023-08-29 &  $+1.201\pm0.032$ & $ -7.443\pm0.472$ & +0.999 &  7.120e-04 \\
2023-08-31 &  $+1.147\pm0.028$ & $ -8.265\pm0.411$ & +0.999 &  5.934e-04 \\
\hline 
\end{tabular}
\\
* Full table is available in its entirety in online version.
\end{table}

\begin{table*}
\centering
\caption{The fit result of LTV of the spectral indices and color behaviors.}
\label{tab:LTV_fit}
\begin{tabular}{lcccc}
\toprule
        Model &  slope &  intercept &  r-value &   p-value\\
\midrule
Time (JD) vs alpha & -4.3e-5$\pm$1.1e-5 & $+107.746\pm+26.937$ & $-0.206$ & 9.086e-05 \\
R-mag vs alpha & $+0.060\pm0.011$ & $  +0.288\pm +0.152$ & $+0.272$ & 1.934e-07 \\
     R-mag  vs B-R & $+0.022\pm0.005$ & $  +0.522\pm +0.071$ & $+0.212$ & 4.045e-05 \\
     R-mag  vs B-I & $+0.039\pm0.007$ & $  +0.814\pm +0.099$ & $+0.269$ & 2.270e-07 \\
     R-mag  vs V-R & $+0.010\pm0.002$ & $  +0.279\pm +0.032$ & $+0.205$ & 5.107e-05 \\
     R-mag  vs V-I & $+0.025\pm0.004$ & $  +0.597\pm +0.059$ & $+0.282$ & 3.105e-08 \\
\bottomrule
\end{tabular}
\end{table*}

\subsection{Cross-correlation Analysis}
In AGN studies, the detection of any time lag between different bands (and the scale of lag if present) provides important clues to the emission region. For this purpose, the Discrete Correlation Function (DCF) is a convenient statistical tool that can be applied to unevenly distributed time series \citep[][and references therein]{2015MNRAS.450..541A}.

The DCF between two discrete data sets ($a_i$, $b_j$) can be computed using the unbinned DCF (UDCF) defined as follows:
\begin{eqnarray}
UDCF_{ij}(\tau) = \frac{(a_i-\bar{a}) (b_j-\bar{b})}{\sqrt{(\sigma_{a}^2-e_{a}^2)(\sigma_{b}^2-e_{b}^2)}}
\end{eqnarray}
Where $\bar{a}$ and $\bar{b}$ represent the mean values of the two time-series data sets, $\sigma_{a,b}$ and $e_{a,b}$ are their standard deviations and errors, respectively. $\Delta t_{ij}=(t_{bj}-t_{ai})$ indicates the time delay between the two data points. To obtain the DCF, the UDCF values are averaged over the interval $\tau - \frac{\Delta\tau}{2} \leq \tau_{ij} \leq \tau + \frac{\Delta\tau}{2}$ as described in the study by \citet{1988ApJ...333..646E}.

\begin{eqnarray}
    DCF(\tau) = \frac{\sum_{k=1}^{m} \text{UDCF}_k}{M}
\end{eqnarray}
The quantity "M" represents the count of pairwise time lag values situated within the specified $\tau$ interval. 

\begin{figure}
\includegraphics[width=\columnwidth]{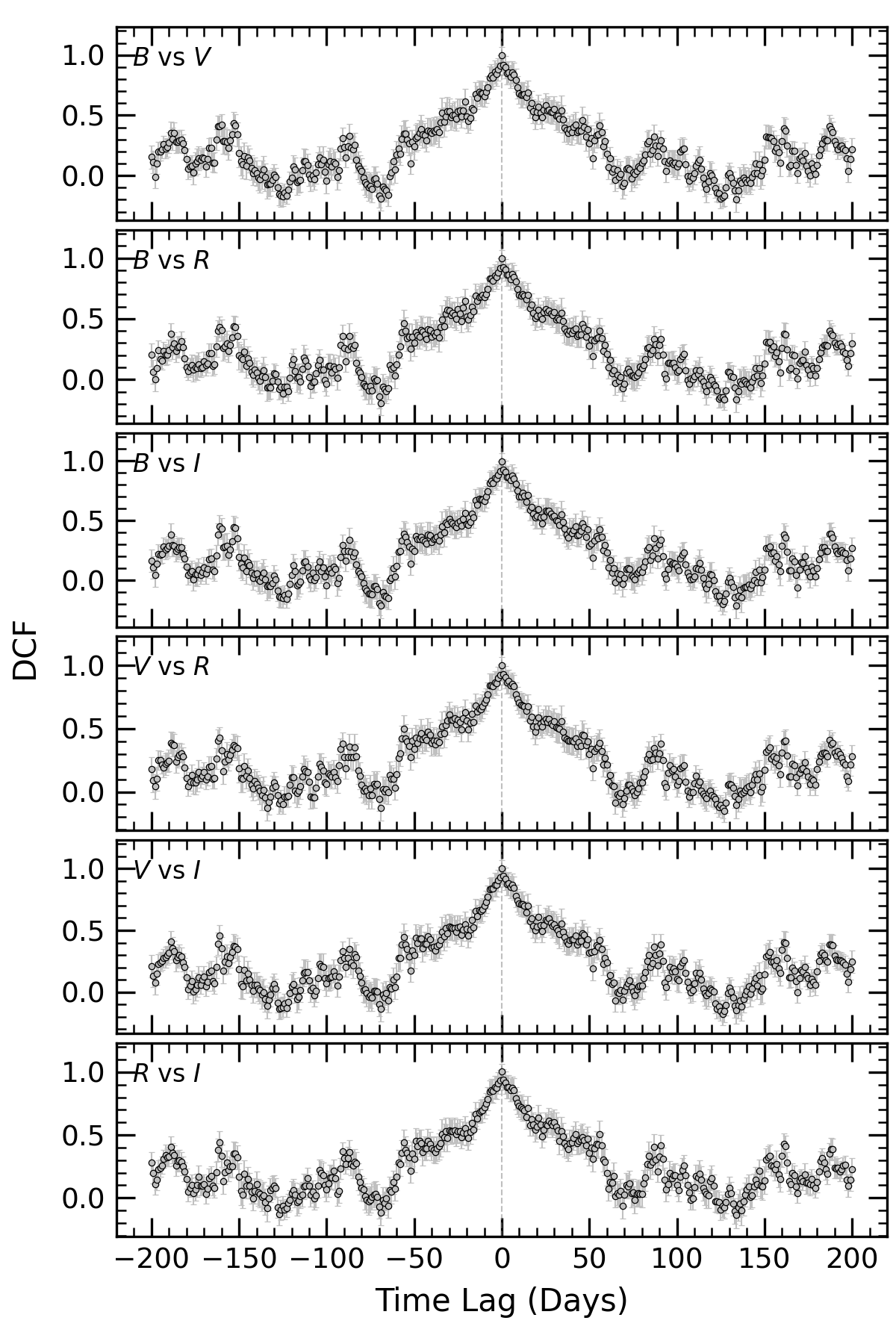}
\caption{Cross-correlation analysis for $B$, $V$, $R$, and $I$ bands using Discrete Correlation Function for the entire monitoring period.}
\label{fig:LTV_dcf}
\end{figure}

For the long-term LCs, we took the weighted mean magnitudes and the mean JD of the nightly binned intranight observations. DCF analysis was applied to each pair combination of the optical BVRI bands for the long-term LCs (Figure \ref{fig:LTV_dcf}). There is a strong correlation between all the bands and no time lag was detected considering a time binning value $\tau$ of 1 day, indicating that the source of the emission is co-spatial.

\subsection{Periodicity Search}

In order to identify any periodicity on the light curve,
the light curves were analyzed using three widely used techniques: Weighted Wavelet Z-transform (WWZ; \cite{1996AJ....112.1709F}) and Lomb-Scargle periodogram (LSP; \cite{1976Ap&SS..39..447L}; \cite{1982ApJ...263..835S}).

The Lomb-Scargle Periodogram (LSP) \citep{1976Ap&SS..39..447L, 1982ApJ...263..835S} is a powerful tool used for detecting periodicity in the time series which can provide clues for interpreting the structure of the central engine. Since the data points in our data set are not evenly distributed, the peaks can be interpreted as quasi-periodic signals in the corresponding time axis.

Although LSP is capable of detecting any periodic pattern in the time series, variations in AGN light curves do not follow a linear groove and they may have more than one periodicity and/or quasi-periodicity. The Weighted Wavelet Z-transform is a method for detecting such oscillations in the light curve.

\begin{figure*}[ht!]
\centering
\includegraphics[width=0.95\textwidth]{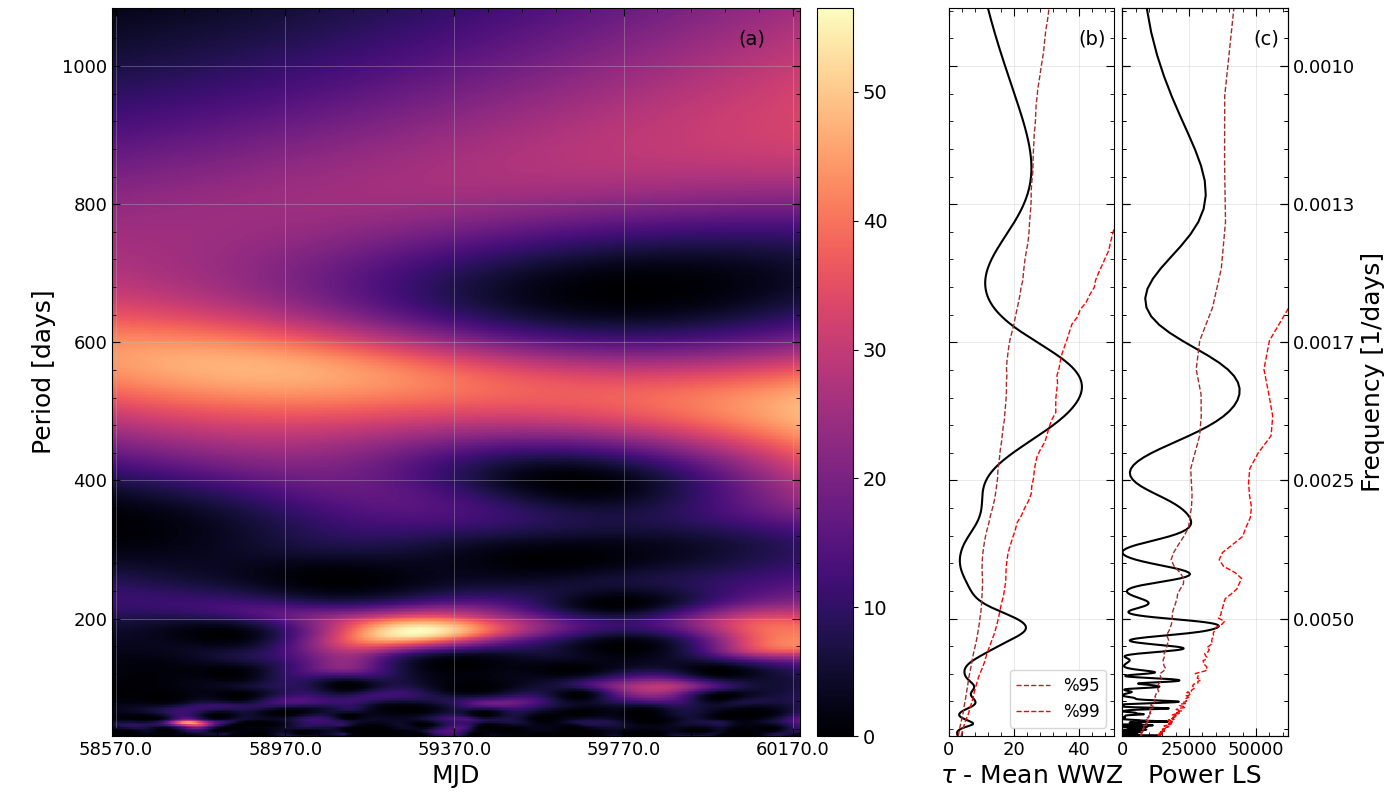}
\caption{Graphs of the WWZ and LS analysis of the long-term optical R-band LC. a) WWZ power spectrum. b) Plot of the WWZ analysis. c) Plot of the LS analysis.}
\label{fig:wwz_ls}
\end{figure*}

The results of the WWZ and LS analysis is shown in the Figure \ref{fig:wwz_ls}. Using the method\footnote{https://github.com/skiehl/lcsim} described in \citet{2022ApJ...926L..35O}, we have simulated 10000 light curves corresponding to the PSD of the blazar LC. We then used WWZ and LS to search for periodicity on the simulated LCs, as we had done on the blazar LC. The significance levels obtained are 95\% and 99\% for each frequency which are shown with purple color in the plots. In the LS plot 188 and 530 days are found within 95\% significance are found in the R-band LC. In the WWZ analysis 186$\pm30$ and 532$\pm76$ days (within 99\% significance) are found in the R-band LC. The results of these analyses in the BVI bands are very similar to those in the R-band. 

\section{Discussion and Conclusions}

{During our observational campaign between March 2019 and August 2023, the blazar was observed quasi-simultaneously for 416 nights in the optical $BVRI$ bands with two optical telescopes, i.e. 1.0\,m RC telescope and 0.6\,m RC robotic telescope in Turkey in both active and quiescent states} The brightness of the source varied between 12.109 and 14.580 in the optical R-band. In the long-term LC, the blazar had a variability amplitude of 256.123\%, 247.084\%, 247.043\%, 230.217 in the BVRI bands respectively. To understand the underlying mechanism of the variability, the flux variability should be analysed together with the color behavior. The blazar did not show a strong color change in the long-term LCs, and the emission in the BVRI bands was strongly correlated without any time lag, which can be interpreted as the source of the emission in these bands being co-spatial. As the spectral window of the BVRI bands is very narrow compared to the entire electromagnetic spectrum of the blazars, it is more difficult to detect any in-correlation. Assuming that the flux comes from a non-thermal radiative process, we have generated nightly spectral indices by fitting a simple power law to the flux in the BVRI bands. The spectral index varies between 0.828 and 1.432 in the long-term LC. When the daily spectral indices are plotted against the R-band magnitude, $\alpha$ becomes steeper as the blazar becomes more active, so to speak, a slight bluer when brighter (BWB) trend can be seen in the spectral energy distributions of the BVRI bands. In addition to the long-term variability, short-term variations (weeks and months time-scale) are clearly visible in the long-term LC.

During the monitoring period we observed the blazar for at least 1 hour in the optical R-band on 21 nights, including 6 quasi-simultaneous observations in the BVRI bands. Statistical tests show that on 12 out of 21 nights the blazar had a significant IDV with a magnitude change of $\approx$0.1 mag and a variability amplitude between 4.41 and 11.23 in the BVRI bands. The variability amplitude of the IDV light curves did not change significantly with state. We have plotted the color-magnitude plots of 6 nights where the BVRI band observations are quasi-simultaneous. Although there are slight slopes in some colors, we cannot claim that there is a clear trend that varies with magnitude on the daily timescale.

To obtain the minimum time-scale, we applied SF to the nightly R-band LCs that have the most significant variability. We found a minimum variability time-scale of $\sim$ 48 min on the LC of 2021-02-10, from which we have calculated the minimum size of the emission region to be $\approx 10^{15} cm$ and the mass of the SMBH to be 5.69 $\times$ $10^8$ $M_\odot$ confirming the result of \cite{2018AJ....156...36K}.

We employed different statistical techniques to detect potential (quasi)-periodicity on the long-term LCs. The WWZ analysis showed that the blazar has quasi-periodicities of 186$\pm30$, 532$\pm76$ days (within 99\% significance) in the optical R-band LC. The LS analyses gives values close to those of the WWZ results, {but the significance of the 586-day QPO falls between 95\% and 99\% levels. According to our simulations the peak in the LSP graph at the $\sim586$ days period is not significant considering the stochastic process. Although we found the similar peak at $523$ days in the WWZ analyses, we cannot claim that this is a reliable QPO, as both analyses would be expected to show a significant peak.}  {We interpret the 186-day QPO in the WWZ power spectrum as a local pattern indicating two outbursts in May 2021 and October 2021.}
We did not detect any other QPO within the limits of significance, confirming the other results of the previous studies mentioned in Section 1.

{Since the contribution from the host galaxy is negligible as a constant component \citep{2022ApJ...928...86G}, it can be assumed that the source of the flux is mainly jet, and hence also the variability. The reason for the long-term variability is suggested by \cite{2013A&A...552A..11R} as geometrical effects. \cite{2013AJ....146..120L} showed that S5 0716+714 has a jet with a changing position angle and can have a helical shape, which leads to a change in the Doppler factor. The change in Doppler factor can cause both short and long term variability and can explain the quasi-chromatic variability of the emission.}

{Another model for the variability is shocks propagating through the jet \citep{1979ApJ...232...34B, 1985ApJ...298..114M}. According to this model, changes in both the flux and color behavior can be caused by relativistic shocks propagating through a jet, and these shocks can interact with different emitting regions with different magnetic and particle compositions \citep{2013A&A...558A..92B}. The variability is interpreted together with the color behavior. The mild color change in the long-term variability and the unpredictable color behavior of the intraday and short-term variability support this model, as the shock front can encounter any such composition in the jet structure. If the emission component is strongly chromatic, it will influence the continuum's color \citep{2015MNRAS.450..541A}. The occasional change in the color behavior can be a result of the inhomogeneity of the plasma both at the base and through the jet. Magnetic reconnection is a mechanism that explains fast variability, namely flares. When the opposing magnetic field lines join and change the magnetic composition, remarkable energy following a power law can be released, as this energy can also be optical \citep{2022ApJ...924...90Z}. The contribution of this process can be considered together with the polarization data. While the source is in the quiescent state, instabilities in the accretion disk, such as hotspots or changes in the accretion rate, can contaminate the LC and cause short-term variability.}

The variability of the blazars is not fully understood. Further simultaneous multi-wavelength and multi-messenger observations together with the theoretical studies would greatly help to fully understand the astrophysical nature of these objects.

\section{Acknowledgments}
We thank TUBITAK National Observatory for partial support in using T60 and T100 telescopes with project numbers 19BT60-1505 and 19AT100-1486, respectively. EE was supported in part by the Scientific Research Project Coordination Unit of Istanbul University (Project ID FDK-2022-39145). This study was supported by Scientific and Technological Research Council of Turkey (TUBITAK) under the Grant Number 121F427.

\software{Astropy \citep{2013A&A...558A..33A, 2018AJ....156..123A, 2022ApJ...935..167A}, ccdproc \citep{2017zndo...1069648C}, 
Numpy \citep{2020Natur.585..357H}, 
Matplotlib \citep{2007CSE.....9...90H}, 
Photutils \citep{2020zndo...4049061B},
WWZ \citep{2023ascl.soft10003K},
lcsim \citep{2023ascl.soft10002K}
}

\paragraph{Data Availability Statement} The data underlying this article are available in the article and in its online material (Figure \ref{fig:IDV_R}; Tables \ref{tab:vartest} and \ref{tab:sed}).

\vspace{5mm}

\pagebreak

\bibliography{new.ms}{}

\begin{thebibliography}{}
\expandafter\ifx\csname natexlab\endcsname\relax\def\natexlab#1{#1}\fi
\providecommand{\url}[1]{\href{#1}{#1}}
\providecommand{\dodoi}[1]{doi:~\href{http://doi.org/#1}{\nolinkurl{#1}}}
\providecommand{\doeprint}[1]{\href{http://ascl.net/#1}{\nolinkurl{http://ascl.net/#1}}}
\providecommand{\doarXiv}[1]{\href{https://arxiv.org/abs/#1}{\nolinkurl{https://arxiv.org/abs/#1}}}

\bibitem[{Abdo {et~al.}(2010)Abdo, Ackermann, Agudo, Ajello, Aller, Aller,
  Angelakis, Arkharov, Axelsson, Bach, Baldini, Ballet, Barbiellini, Bastieri,
  Baughman, Bechtol, Bellazzini, Benitez, Berdyugin, Berenji, Blandford, Bloom,
  Boettcher, Bonamente, Borgland, Bregeon, Brez, Brigida, Bruel, Burnett,
  Burrows, Buson, Caliandro, Calzoletti, Cameron, Capalbi, Caraveo, Carosati,
  Casandjian, Cavazzuti, Cecchi, Çelik, Charles, Chaty, Chekhtman, Chen,
  Chiang, Chincarini, Ciprini, Claus, Cohen-Tanugi, Colafrancesco, Cominsky,
  Conrad, Costamante, Cutini, D'ammando, Deitrick, D'Elia, Dermer, de~Angelis,
  de~Palma, Digel, Donnarumma, do~Couto~e Silva, Drell, Dubois, Dultzin,
  Dumora, Falcone, Farnier, Favuzzi, Fegan, Focke, Forné, Fortin, Frailis,
  Fuhrmann, Fukazawa, Funk, Fusco, Gómez, Gargano, Gasparrini, Gehrels,
  Germani, Giebels, Giglietto, Giommi, Giordano, Giuliani, Glanzman, Godfrey,
  Grenier, Gronwall, Grove, Guillemot, Guiriec, Gurwell, Hadasch, Hanabata,
  Harding, Hayashida, Hays, Healey, Heidt, Hiriart, Horan, Hoversten, Hughes,
  Itoh, Jackson, Jóhannesson, Johnson, Johnson, Jorstad, Kadler, Kamae,
  Katagiri, Kataoka, Kawai, Kennea, Kerr, Kimeridze, Knödlseder, Kocian,
  Kopatskaya, Koptelova, Konstantinova, Kovalev, Kovalev, Kurtanidze, Kuss,
  Lande, Larionov, Latronico, Leto, Lindfors, Longo, Loparco, Lott, Lovellette,
  Lubrano, Madejski, Makeev, Marchegiani, Marscher, Marshall, Max-Moerbeck,
  Mazziotta, McConville, McEnery, Meurer, Michelson, Mitthumsiri, Mizuno,
  Moiseev, Monte, Monzani, Morselli, Moskalenko, Murgia, Nestoras, Nilsson,
  Nizhelsky, Nolan, Norris, Nuss, Ohsugi, Ojha, Omodei, Orlando, Ormes,
  Osborne, Ozaki, Pacciani, Padovani, Pagani, Page, Paneque, Panetta, Parent,
  Pasanen, Pavlidou, Pelassa, Pepe, Perri, Pesce-Rollins, Piranomonte, Piron,
  Pittori, Porter, Puccetti, Rahoui, Rainò, Raiteri, Rando, Razzano, Reimer,
  Reimer, Reposeur, Richards, Ritz, Rochester, Rodriguez, Romani, Ros, Roth,
  Roustazadeh, Ryde, Sadrozinski, Sadun, Sanchez, Sander, Parkinson, Scargle,
  Sellerholm, Sgrò, Shaw, Sigua, Siskind, Smith, Smith, Spandre, Spinelli,
  Starck, Stevenson, Stratta, Strickman, Suson, Tajima, Takahashi, Takahashi,
  Takalo, Tanaka, Thayer, Thayer, Thompson, Tibaldo, Torres, Tosti, Tramacere,
  Uchiyama, Usher, Vasileiou, Verrecchia, Vilchez, Villata, Vitale, Waite,
  Wang, Winer, Wood, Ylinen, Zensus, Zhekanis, \& Ziegler}]{Abdo_2010}
Abdo, A.~A., Ackermann, M., Agudo, I., {et~al.} 2010, The Astrophysical
  Journal, 716, 30, \dodoi{10.1088/0004-637X/716/1/30}

\bibitem[{{Abramowicz} \& {Nobili}(1982)}]{1982Natur.300..506A}
{Abramowicz}, M.~A., \& {Nobili}, L. 1982, \nat, 300, 506,
  \dodoi{10.1038/300506a0}

\bibitem[{{Agarwal} \& {Gupta}(2015)}]{2015MNRAS.450..541A}
{Agarwal}, A., \& {Gupta}, A.~C. 2015, \mnras, 450, 541,
  \dodoi{10.1093/mnras/stv625}

\bibitem[{{Agarwal} {et~al.}(2022){Agarwal}, {Pandey}, {{\"O}zd{\"o}nmez},
  {Ege}, {Kumar Das}, \& {Karakulak}}]{2022ApJ...933...42A}
{Agarwal}, A., {Pandey}, A., {{\"O}zd{\"o}nmez}, A., {et~al.} 2022, \apj, 933,
  42, \dodoi{10.3847/1538-4357/ac6cef}

\bibitem[{{Agarwal} {et~al.}(2016){Agarwal}, {Gupta}, {Bachev}, {Strigachev},
  {Semkov}, {Wiita}, {Fan}, {Pandey}, {Boeva}, \&
  {Spassov}}]{2016MNRAS.455..680A}
{Agarwal}, A., {Gupta}, A.~C., {Bachev}, R., {et~al.} 2016, \mnras, 455, 680,
  \dodoi{10.1093/mnras/stv2345}

\bibitem[{{Anderhub} {et~al.}(2009){Anderhub}, {Antonelli}, {Antoranz},
  {Backes}, {Baixeras}, {Balestra}, {Barrio}, {Bastieri}, {Becerra
  Gonz{\'a}lez}, {Becker}, {Bednarek}, {Berdyugin}, {Berger}, {Bernardini},
  {Biland}, {Bock}, {Bonnoli}, {Bordas}, {Borla Tridon}, {Bosch-Ramon}, {Bose},
  {Braun}, {Bretz}, {Britzger}, {Camara}, {Carmona}, {Carosi}, {Colin},
  {Commichau}, {Contreras}, {Cortina}, {Costado}, {Covino}, {Dazzi}, {De
  Angelis}, {de Cea del Pozo}, {De los Reyes}, {De Lotto}, {De Maria}, {De
  Sabata}, {Delgado Mendez}, {Dom{\'\i}nguez}, {Dominis Prester}, {Dorner},
  {Doro}, {Elsaesser}, {Errando}, {Ferenc}, {Fern{\'a}ndez}, {Firpo},
  {Fonseca}, {Font}, {Galante}, {Garc{\'\i}a L{\'o}pez}, {Garczarczyk}, {Gaug},
  {Godinovic}, {Goebel}, {Hadasch}, {Herrero}, {Hildebrand},
  {H{\"o}hne-M{\"o}nch}, {Hose}, {Hrupec}, {Hsu}, {Jogler}, {Klepser},
  {Kranich}, {La Barbera}, {Laille}, {Leonardo}, {Lindfors}, {Lombardi},
  {Longo}, {L{\'o}pez}, {Lorenz}, {Majumdar}, {Maneva}, {Mankuzhiyil},
  {Mannheim}, {Maraschi}, {Mariotti}, {Mart{\'\i}nez}, {Mazin}, {Meucci},
  {Miranda}, {Mirzoyan}, {Miyamoto}, {Mold{\'o}n}, {Moles}, {Moralejo},
  {Nieto}, {Nilsson}, {Ninkovic}, {Orito}, {Oya}, {Paoletti}, {Paredes},
  {Pasanen}, {Pascoli}, {Pauss}, {Pegna}, {Perez-Torres}, {Persic}, {Peruzzo},
  {Prada}, {Prandini}, {Puchades}, {Puljak}, {Reichardt}, {Rhode}, {Rib{\'o}},
  {Rico}, {Rissi}, {Robert}, {R{\"u}gamer}, {Saggion}, {Sainio}, {Saito},
  {Salvati}, {S{\'a}nchez-Conde}, {Satalecka}, {Scalzotto}, {Scapin},
  {Schweizer}, {Shayduk}, {Shore}, {Sierpowska-Bartosik}, {Sillanp{\"a}{\"a}},
  {Sitarek}, {Sobczynska}, {Spanier}, {Spiro}, {Stamerra}, {Stark}, {Suric},
  {Takalo}, {Tavecchio}, {Temnikov}, {Tescaro}, {Teshima}, {Torres}, {Turini},
  {Vankov}, {Wagner}, {Villforth}, {Zabalza}, {Zandanel}, {Zanin}, \&
  {Zapatero}}]{2009ApJ...704L.129A}
{Anderhub}, H., {Antonelli}, L.~A., {Antoranz}, P., {et~al.} 2009, \apjl, 704,
  L129, \dodoi{10.1088/0004-637X/704/2/L129}

\bibitem[{{Astropy Collaboration} {et~al.}(2013){Astropy Collaboration},
  {Robitaille}, {Tollerud}, {Greenfield}, {Droettboom}, {Bray}, {Aldcroft},
  {Davis}, {Ginsburg}, {Price-Whelan}, {Kerzendorf}, {Conley}, {Crighton},
  {Barbary}, {Muna}, {Ferguson}, {Grollier}, {Parikh}, {Nair}, {Unther},
  {Deil}, {Woillez}, {Conseil}, {Kramer}, {Turner}, {Singer}, {Fox}, {Weaver},
  {Zabalza}, {Edwards}, {Azalee Bostroem}, {Burke}, {Casey}, {Crawford},
  {Dencheva}, {Ely}, {Jenness}, {Labrie}, {Lim}, {Pierfederici}, {Pontzen},
  {Ptak}, {Refsdal}, {Servillat}, \& {Streicher}}]{2013A&A...558A..33A}
{Astropy Collaboration}, {Robitaille}, T.~P., {Tollerud}, E.~J., {et~al.} 2013,
  \aap, 558, A33, \dodoi{10.1051/0004-6361/201322068}

\bibitem[{{Astropy Collaboration} {et~al.}(2018){Astropy Collaboration},
  {Price-Whelan}, {Sip{\H{o}}cz}, {G{\"u}nther}, {Lim}, {Crawford}, {Conseil},
  {Shupe}, {Craig}, {Dencheva}, {Ginsburg}, {VanderPlas}, {Bradley},
  {P{\'e}rez-Su{\'a}rez}, {de Val-Borro}, {Aldcroft}, {Cruz}, {Robitaille},
  {Tollerud}, {Ardelean}, {Babej}, {Bach}, {Bachetti}, {Bakanov}, {Bamford},
  {Barentsen}, {Barmby}, {Baumbach}, {Berry}, {Biscani}, {Boquien}, {Bostroem},
  {Bouma}, {Brammer}, {Bray}, {Breytenbach}, {Buddelmeijer}, {Burke},
  {Calderone}, {Cano Rodr{\'\i}guez}, {Cara}, {Cardoso}, {Cheedella}, {Copin},
  {Corrales}, {Crichton}, {D'Avella}, {Deil}, {Depagne}, {Dietrich}, {Donath},
  {Droettboom}, {Earl}, {Erben}, {Fabbro}, {Ferreira}, {Finethy}, {Fox},
  {Garrison}, {Gibbons}, {Goldstein}, {Gommers}, {Greco}, {Greenfield},
  {Groener}, {Grollier}, {Hagen}, {Hirst}, {Homeier}, {Horton}, {Hosseinzadeh},
  {Hu}, {Hunkeler}, {Ivezi{\'c}}, {Jain}, {Jenness}, {Kanarek}, {Kendrew},
  {Kern}, {Kerzendorf}, {Khvalko}, {King}, {Kirkby}, {Kulkarni}, {Kumar},
  {Lee}, {Lenz}, {Littlefair}, {Ma}, {Macleod}, {Mastropietro}, {McCully},
  {Montagnac}, {Morris}, {Mueller}, {Mumford}, {Muna}, {Murphy}, {Nelson},
  {Nguyen}, {Ninan}, {N{\"o}the}, {Ogaz}, {Oh}, {Parejko}, {Parley}, {Pascual},
  {Patil}, {Patil}, {Plunkett}, {Prochaska}, {Rastogi}, {Reddy Janga},
  {Sabater}, {Sakurikar}, {Seifert}, {Sherbert}, {Sherwood-Taylor}, {Shih},
  {Sick}, {Silbiger}, {Singanamalla}, {Singer}, {Sladen}, {Sooley},
  {Sornarajah}, {Streicher}, {Teuben}, {Thomas}, {Tremblay}, {Turner},
  {Terr{\'o}n}, {van Kerkwijk}, {de la Vega}, {Watkins}, {Weaver}, {Whitmore},
  {Woillez}, {Zabalza}, \& {Astropy Contributors}}]{2018AJ....156..123A}
{Astropy Collaboration}, {Price-Whelan}, A.~M., {Sip{\H{o}}cz}, B.~M., {et~al.}
  2018, \aj, 156, 123, \dodoi{10.3847/1538-3881/aabc4f}

\bibitem[{{Astropy Collaboration} {et~al.}(2022){Astropy Collaboration},
  {Price-Whelan}, {Lim}, {Earl}, {Starkman}, {Bradley}, {Shupe}, {Patil},
  {Corrales}, {Brasseur}, {N{\"o}the}, {Donath}, {Tollerud}, {Morris},
  {Ginsburg}, {Vaher}, {Weaver}, {Tocknell}, {Jamieson}, {van Kerkwijk},
  {Robitaille}, {Merry}, {Bachetti}, {G{\"u}nther}, {Aldcroft},
  {Alvarado-Montes}, {Archibald}, {B{\'o}di}, {Bapat}, {Barentsen},
  {Baz{\'a}n}, {Biswas}, {Boquien}, {Burke}, {Cara}, {Cara}, {Conroy},
  {Conseil}, {Craig}, {Cross}, {Cruz}, {D'Eugenio}, {Dencheva}, {Devillepoix},
  {Dietrich}, {Eigenbrot}, {Erben}, {Ferreira}, {Foreman-Mackey}, {Fox},
  {Freij}, {Garg}, {Geda}, {Glattly}, {Gondhalekar}, {Gordon}, {Grant},
  {Greenfield}, {Groener}, {Guest}, {Gurovich}, {Handberg}, {Hart},
  {Hatfield-Dodds}, {Homeier}, {Hosseinzadeh}, {Jenness}, {Jones}, {Joseph},
  {Kalmbach}, {Karamehmetoglu}, {Ka{\l}uszy{\'n}ski}, {Kelley}, {Kern},
  {Kerzendorf}, {Koch}, {Kulumani}, {Lee}, {Ly}, {Ma}, {MacBride}, {Maljaars},
  {Muna}, {Murphy}, {Norman}, {O'Steen}, {Oman}, {Pacifici}, {Pascual},
  {Pascual-Granado}, {Patil}, {Perren}, {Pickering}, {Rastogi}, {Roulston},
  {Ryan}, {Rykoff}, {Sabater}, {Sakurikar}, {Salgado}, {Sanghi}, {Saunders},
  {Savchenko}, {Schwardt}, {Seifert-Eckert}, {Shih}, {Jain}, {Shukla}, {Sick},
  {Simpson}, {Singanamalla}, {Singer}, {Singhal}, {Sinha}, {Sip{\H{o}}cz},
  {Spitler}, {Stansby}, {Streicher}, {{\v{S}}umak}, {Swinbank}, {Taranu},
  {Tewary}, {Tremblay}, {de Val-Borro}, {Van Kooten}, {Vasovi{\'c}}, {Verma},
  {de Miranda Cardoso}, {Williams}, {Wilson}, {Winkel}, {Wood-Vasey}, {Xue},
  {Yoachim}, {Zhang}, {Zonca}, \& {Astropy Project
  Contributors}}]{2022ApJ...935..167A}
{Astropy Collaboration}, {Price-Whelan}, A.~M., {Lim}, P.~L., {et~al.} 2022,
  \apj, 935, 167, \dodoi{10.3847/1538-4357/ac7c74}

\bibitem[{{Bessell} {et~al.}(1998){Bessell}, {Castelli}, \&
  {Plez}}]{1998A&A...333..231B}
{Bessell}, M.~S., {Castelli}, F., \& {Plez}, B. 1998, \aap, 333, 231

\bibitem[{{Bhatta} {et~al.}(2013){Bhatta}, {Webb}, {Hollingsworth}, {Dhalla},
  {Khanuja}, {Bachev}, {Blinov}, {B{\"o}ttcher}, {Bravo Calle}, {Calcidese},
  {Capezzali}, {Carosati}, {Chigladze}, {Collins}, {Coloma}, {Efimov}, {Gupta},
  {Hu}, {Kurtanidze}, {Lamerato}, {Larionov}, {Lee}, {Lindfors}, {Murphy},
  {Nilsson}, {Ohlert}, {Oksanen}, {P{\"a}{\"a}kk{\"o}nen}, {Pollock}, {Rani},
  {Reinthal}, {Rodriguez}, {Ros}, {Roustazadeh}, {Sagar}, {Sanchez}, {Shastri},
  {Sillanp{\"a}{\"a}}, {Strigachev}, {Takalo}, {Vennes}, {Villata},
  {Villforth}, {Wu}, \& {Zhou}}]{2013A&A...558A..92B}
{Bhatta}, G., {Webb}, J.~R., {Hollingsworth}, H., {et~al.} 2013, \aap, 558,
  A92, \dodoi{10.1051/0004-6361/201220236}

\bibitem[{{Bhatta} {et~al.}(2016){Bhatta}, {Stawarz}, {Ostrowski}, {Markowitz},
  {Akitaya}, {Arkharov}, {Bachev}, {Ben{\'\i}tez}, {Borman}, {Carosati},
  {Cason}, {Chanishvili}, {Damljanovic}, {Dhalla}, {Frasca}, {Hiriart}, {Hu},
  {Itoh}, {Jableka}, {Jorstad}, {Jovanovic}, {Kawabata}, {Klimanov},
  {Kurtanidze}, {Larionov}, {Laurence}, {Leto}, {Marscher}, {Moody},
  {Moritani}, {Ohlert}, {Di Paola}, {Raiteri}, {Rizzi}, {Sadun}, {Sasada},
  {Sergeev}, {Strigachev}, {Takaki}, {Troitsky}, {Ui}, {Villata}, {Vince},
  {Webb}, {Yoshida}, \& {Zola}}]{2016ApJ...831...92B}
{Bhatta}, G., {Stawarz}, {\L}., {Ostrowski}, M., {et~al.} 2016, \apj, 831, 92,
  \dodoi{10.3847/0004-637X/831/1/92}

\bibitem[{{Blandford} \& {K{\"o}nigl}(1979)}]{1979ApJ...232...34B}
{Blandford}, R.~D., \& {K{\"o}nigl}, A. 1979, \apj, 232, 34,
  \dodoi{10.1086/157262}

\bibitem[{{Bradley} {et~al.}(2020){Bradley}, {Sip{\H{o}}cz}, {Robitaille},
  {Tollerud}, {Vin{\'\i}cius}, {Deil}, {Barbary}, {Wilson}, {Busko},
  {G{\"u}nther}, {Cara}, {Conseil}, {Bostroem}, {Droettboom}, {Bray}, {Andersen
  Bratholm}, {Lim}, {Barentsen}, {Craig}, {Pascual}, {Perren}, {Greco},
  {Donath}, {De Val-Borro}, {Kerzendorf}, {Bach}, {Weaver}, {D'Eugenio},
  {Souchereau}, \& {Ferreira}}]{2020zndo...4049061B}
{Bradley}, L., {Sip{\H{o}}cz}, B., {Robitaille}, T., {et~al.} 2020,
  {astropy/photutils: 1.0.1}, 1.0.1,  Zenodo, \dodoi{10.5281/zenodo.4049061}

\bibitem[{{Carrasco} {et~al.}(2019){Carrasco}, {Escobedo}, {Recillas},
  {Porras}, {Chavushyan}, \& {Patino-Alvarez}}]{2019ATel13359....1C}
{Carrasco}, L., {Escobedo}, G., {Recillas}, E., {et~al.} 2019, The Astronomer's
  Telegram, 13359, 1

\bibitem[{{Chandra} {et~al.}(2011){Chandra}, {Baliyan}, {Ganesh}, \&
  {Joshi}}]{2011ApJ...731..118C}
{Chandra}, S., {Baliyan}, K.~S., {Ganesh}, S., \& {Joshi}, U.~C. 2011, \apj,
  731, 118, \dodoi{10.1088/0004-637X/731/2/118}

\bibitem[{{Chandra} {et~al.}(2015){Chandra}, {Zhang}, {Kushwaha}, {Singh},
  {Bottcher}, {Kaur}, \& {Baliyan}}]{2015ApJ...809..130C}
{Chandra}, S., {Zhang}, H., {Kushwaha}, P., {et~al.} 2015, \apj, 809, 130,
  \dodoi{10.1088/0004-637X/809/2/130}

\bibitem[{Chatterjee {et~al.}(2018)Chatterjee, Roychowdhury, Chandra, \&
  Sinha}]{Chatterjee_2018}
Chatterjee, R., Roychowdhury, A., Chandra, S., \& Sinha, A. 2018, The
  Astrophysical Journal Letters, 859, L21, \dodoi{10.3847/2041-8213/aac48a}

\bibitem[{{Chen} {et~al.}(2022){Chen}, {Yi}, {Gong}, {Yang}, {Chen}, {Chang},
  \& {Mao}}]{2022ApJ...938....8C}
{Chen}, J., {Yi}, T., {Gong}, Y., {et~al.} 2022, \apj, 938, 8,
  \dodoi{10.3847/1538-4357/ac91c3}

\bibitem[{{Craig} {et~al.}(2017){Craig}, {Crawford}, {Seifert}, {Robitaille},
  {Sipocz}, {Walawender}, {Vin{\'\i}cius}, {Ninan}, {Droettboom}, {Youn},
  {Tollerud}, {Bray}, {Walkerna22}, {Reddy Janga}, {Stottsco}, {G{\"u}nther},
  {Rol}, {Bach}, {Bradley}, {Deil}, {Price-Whelan}, {Barbary}, {Horton},
  {Schoenell}, {Nathan}, {Gasdia}, {Nelson}, \&
  {Streicher}}]{2017zndo...1069648C}
{Craig}, M., {Crawford}, S., {Seifert}, M., {et~al.} 2017, {astropy/ccdproc:
  v1.3.0.post1}, v1.3.0.post1,  Zenodo, \dodoi{10.5281/zenodo.1069648}

\bibitem[{{Dai} {et~al.}(2015){Dai}, {Zeng}, {Jiang}, {Fan}, {Hu}, {Zhang},
  {Yang}, {Yan}, {Wang}, \& {Zhang}}]{2015ApJS..218...18D}
{Dai}, B.-z., {Zeng}, W., {Jiang}, Z.-j., {et~al.} 2015, \apjs, 218, 18,
  \dodoi{10.1088/0067-0049/218/2/18}

\bibitem[{{de Diego}(2014)}]{2014AJ....148...93D}
{de Diego}, J.~A. 2014, \aj, 148, 93, \dodoi{10.1088/0004-6256/148/5/93}

\bibitem[{{de Diego} {et~al.}(2015){de Diego}, {Polednikova}, {Bongiovanni},
  {P{\'e}rez Garc{\'\i}a}, {De Leo}, {Verdugo}, \&
  {Cepa}}]{2015AJ....150...44D}
{de Diego}, J.~A., {Polednikova}, J., {Bongiovanni}, A., {et~al.} 2015, \aj,
  150, 44, \dodoi{10.1088/0004-6256/150/2/44}

\bibitem[{{Edelson} \& {Krolik}(1988)}]{1988ApJ...333..646E}
{Edelson}, R.~A., \& {Krolik}, J.~H. 1988, \apj, 333, 646,
  \dodoi{10.1086/166773}

\bibitem[{{Elliot} \& {Shapiro}(1974)}]{1974ApJ...192L...3E}
{Elliot}, J.~L., \& {Shapiro}, S.~L. 1974, \apjl, 192, L3,
  \dodoi{10.1086/181575}

\bibitem[{{Fan} {et~al.}(2016){Fan}, {Yang}, {Liu}, {Luo}, {Lin}, {Yuan},
  {Xiao}, {Zhou}, {Hua}, \& {Pei}}]{2016ApJS..226...20F}
{Fan}, J.~H., {Yang}, J.~H., {Liu}, Y., {et~al.} 2016, \apjs, 226, 20,
  \dodoi{10.3847/0067-0049/226/2/20}

\bibitem[{{Foster}(1996)}]{1996AJ....112.1709F}
{Foster}, G. 1996, \aj, 112, 1709, \dodoi{10.1086/118137}

\bibitem[{{Gallo} {et~al.}(2018){Gallo}, {Blue}, {Grupe}, {Komossa}, \&
  {Wilkins}}]{2018MNRAS.478.2557G}
{Gallo}, L.~C., {Blue}, D.~M., {Grupe}, D., {Komossa}, S., \& {Wilkins}, D.~R.
  2018, \mnras, 478, 2557, \dodoi{10.1093/mnras/sty1134}

\bibitem[{{Gaur} {et~al.}(2015){Gaur}, {Gupta}, {Bachev}, {Strigachev},
  {Semkov}, {B{\"o}ttcher}, {Wiita}, {de Diego}, {Gu}, {Guo}, {Joshi}, {Mihov},
  {Palma}, {Peneva}, {Rajasingam}, \& {Slavcheva-Mihova}}]{2015MNRAS.452.4263G}
{Gaur}, H., {Gupta}, A.~C., {Bachev}, R., {et~al.} 2015, \mnras, 452, 4263,
  \dodoi{10.1093/mnras/stv1556}

\bibitem[{{Ghisellini} {et~al.}(1997){Ghisellini}, {Villata}, {Raiteri},
  {Bosio}, {de Francesco}, {Latini}, {Maesano}, {Massaro}, {Montagni}, {Nesci},
  {Tosti}, {Fiorucci}, {Pian}, {Maraschi}, {Treves}, {Comastri}, \&
  {Mignoli}}]{1997A&A...327...61G}
{Ghisellini}, G., {Villata}, M., {Raiteri}, C.~M., {et~al.} 1997, \aap, 327,
  61, \dodoi{10.48550/arXiv.astro-ph/9706254}

\bibitem[{{Gorbachev} {et~al.}(2022){Gorbachev}, {Butuzova}, {Sergeev},
  {Nazarov}, \& {Zhovtan}}]{2022ApJ...928...86G}
{Gorbachev}, M.~A., {Butuzova}, M.~S., {Sergeev}, S.~G., {Nazarov}, S.~V., \&
  {Zhovtan}, A.~V. 2022, \apj, 928, 86, \dodoi{10.3847/1538-4357/ac4fc3}

\bibitem[{{Gupta} {et~al.}(2008{\natexlab{a}}){Gupta}, {Fan}, {Bai}, \&
  {Wagner}}]{2008AJ....135.1384G}
{Gupta}, A.~C., {Fan}, J.~H., {Bai}, J.~M., \& {Wagner}, S.~J.
  2008{\natexlab{a}}, \aj, 135, 1384, \dodoi{10.1088/0004-6256/135/4/1384}

\bibitem[{{Gupta} \& {Joshi}(2005)}]{2005A&A...440..855G}
{Gupta}, A.~C., \& {Joshi}, U.~C. 2005, \aap, 440, 855,
  \dodoi{10.1051/0004-6361:20042370}

\bibitem[{{Gupta} {et~al.}(2008{\natexlab{b}}){Gupta}, {Cha}, {Lee}, {Jin},
  {Pak}, {Cho}, {Moon}, {Park}, {Yuk}, {Nam}, \&
  {Kyeong}}]{2008AJ....136.2359G}
{Gupta}, A.~C., {Cha}, S.-M., {Lee}, S., {et~al.} 2008{\natexlab{b}}, \aj, 136,
  2359, \dodoi{10.1088/0004-6256/136/6/2359}

\bibitem[{{Gupta} {et~al.}(2012){Gupta}, {Krichbaum}, {Wiita}, {Rani},
  {Sokolovsky}, {Mohan}, {Mangalam}, {Marchili}, {Fuhrmann}, {Agudo}, {Bach},
  {Bachev}, {B{\"o}ttcher}, {Gabanyi}, {Gaur}, {Hawkins}, {Kimeridze},
  {Kurtanidze}, {Kurtanidze}, {Lee}, {Liu}, {McBreen}, {Nesci}, {Nestoras},
  {Nikolashvili}, {Ohlert}, {Palma}, {Peneva}, {Pursimo}, {Semkov},
  {Strigachev}, {Webb}, {Wiesemeyer}, \& {Zensus}}]{2012MNRAS.425.1357G}
{Gupta}, A.~C., {Krichbaum}, T.~P., {Wiita}, P.~J., {et~al.} 2012, \mnras, 425,
  1357, \dodoi{10.1111/j.1365-2966.2012.21550.x}

\bibitem[{{Haiyan} {et~al.}(2023){Haiyan}, {Xiefei}, {Xiaopan}, {Na}, {Haitao},
  {Yuhui}, {Li}, \& {Yan}}]{2023Ap&SS.368...88H}
{Haiyan}, Y., {Xiefei}, S., {Xiaopan}, L., {et~al.} 2023, \apss, 368, 88,
  \dodoi{10.1007/s10509-023-04247-6}

\bibitem[{{Harris} {et~al.}(2020){Harris}, {Millman}, {van der Walt},
  {Gommers}, {Virtanen}, {Cournapeau}, {Wieser}, {Taylor}, {Berg}, {Smith},
  {Kern}, {Picus}, {Hoyer}, {van Kerkwijk}, {Brett}, {Haldane}, {del R{\'\i}o},
  {Wiebe}, {Peterson}, {G{\'e}rard-Marchant}, {Sheppard}, {Reddy}, {Weckesser},
  {Abbasi}, {Gohlke}, \& {Oliphant}}]{2020Natur.585..357H}
{Harris}, C.~R., {Millman}, K.~J., {van der Walt}, S.~J., {et~al.} 2020, \nat,
  585, 357, \dodoi{10.1038/s41586-020-2649-2}

\bibitem[{{Heidt} \& {Wagner}(1996)}]{1996A&A...305...42H}
{Heidt}, J., \& {Wagner}, S.~J. 1996, \aap, 305, 42,
  \dodoi{10.48550/arXiv.astro-ph/9506032}

\bibitem[{{Hong} {et~al.}(2017){Hong}, {Xiong}, \& {Bai}}]{2017AJ....154...42H}
{Hong}, S., {Xiong}, D., \& {Bai}, J. 2017, \aj, 154, 42,
  \dodoi{10.3847/1538-3881/aa799a}

\bibitem[{{Hong} {et~al.}(2018){Hong}, {Xiong}, \& {Bai}}]{2018AJ....155...31H}
---. 2018, \aj, 155, 31, \dodoi{10.3847/1538-3881/aa9d89}

\bibitem[{{Hunter}(2007)}]{2007CSE.....9...90H}
{Hunter}, J.~D. 2007, Computing in Science and Engineering, 9, 90,
  \dodoi{10.1109/MCSE.2007.55}

\bibitem[{{Kaur} {et~al.}(2018){Kaur}, {Baliyan}, {Chandra}, {Sameer}, \&
  {Ganesh}}]{2018AJ....156...36K}
{Kaur}, N., {Baliyan}, K.~S., {Chandra}, S., {Sameer}, \& {Ganesh}, S. 2018,
  \aj, 156, 36, \dodoi{10.3847/1538-3881/aac5e4}

\bibitem[{{Kiehlmann}(2023)}]{2023ascl.soft10002K}
{Kiehlmann}, S. 2023, {lcsim: Light curve simulation code}, Astrophysics Source
  Code Library, record ascl:2310.002.
\newblock \doeprint{2310.002}

\bibitem[{{Kiehlmann} {et~al.}(2023){Kiehlmann}, {Max-Moerbeck}, \&
  {King}}]{2023ascl.soft10003K}
{Kiehlmann}, S., {Max-Moerbeck}, W., \& {King}, O. 2023, {wwz: Weighted wavelet
  z-transform code}, Astrophysics Source Code Library, record ascl:2310.003

\bibitem[{{Krawczynski}(2004)}]{2004NewAR..48..367K}
{Krawczynski}, H. 2004, \nar, 48, 367, \dodoi{10.1016/j.newar.2003.12.008}

\bibitem[{{Kudryavtseva} {et~al.}(2011){Kudryavtseva}, {Britzen}, {Witzel},
  {Ros}, {Karouzos}, {Aller}, {Aller}, {Ter{\"a}sranta}, {Eckart}, \&
  {Zensus}}]{2011A&A...526A..51K}
{Kudryavtseva}, N.~A., {Britzen}, S., {Witzel}, A., {et~al.} 2011, \aap, 526,
  A51, \dodoi{10.1051/0004-6361/201014968}

\bibitem[{{Kuehr} {et~al.}(1981){Kuehr}, {Pauliny-Toth}, {Witzel}, \&
  {Schmidt}}]{1981AJ.....86..854K}
{Kuehr}, H., {Pauliny-Toth}, I.~I.~K., {Witzel}, A., \& {Schmidt}, J. 1981,
  \aj, 86, 854, \dodoi{10.1086/112957}

\bibitem[{Li {et~al.}(2023)Li, Yang, Cai, Lähteenmäki, Tornikoski, Tammi,
  Suutarinen, Yang, Luo, \& Wang}]{Li_2023}
Li, X.-P., Yang, H.-Y., Cai, Y., {et~al.} 2023, The Astrophysical Journal, 943,
  157, \dodoi{10.3847/1538-4357/acae8c}

\bibitem[{{Liao} {et~al.}(2014){Liao}, {Bai}, {Liu}, {Weng}, {Chen}, \&
  {Li}}]{2014ApJ...783...83L}
{Liao}, N.~H., {Bai}, J.~M., {Liu}, H.~T., {et~al.} 2014, \apj, 783, 83,
  \dodoi{10.1088/0004-637X/783/2/83}

\bibitem[{Liao {et~al.}(2014)Liao, Bai, Liu, Weng, Chen, \& Li}]{Liao_2014}
Liao, N.~H., Bai, J.~M., Liu, H.~T., {et~al.} 2014, The Astrophysical Journal,
  783, 83, \dodoi{10.1088/0004-637X/783/2/83}

\bibitem[{{Lister} {et~al.}(2013){Lister}, {Aller}, {Aller}, {Homan},
  {Kellermann}, {Kovalev}, {Pushkarev}, {Richards}, {Ros}, \&
  {Savolainen}}]{2013AJ....146..120L}
{Lister}, M.~L., {Aller}, M.~F., {Aller}, H.~D., {et~al.} 2013, \aj, 146, 120,
  \dodoi{10.1088/0004-6256/146/5/120}

\bibitem[{{Lomb}(1976)}]{1976Ap&SS..39..447L}
{Lomb}, N.~R. 1976, \apss, 39, 447, \dodoi{10.1007/BF00648343}

\bibitem[{{Marscher} \& {Gear}(1985)}]{1985ApJ...298..114M}
{Marscher}, A.~P., \& {Gear}, W.~K. 1985, \apj, 298, 114,
  \dodoi{10.1086/163592}

\bibitem[{{Marscher} \& {Travis}(1996)}]{1996A&AS..120C.537M}
{Marscher}, A.~P., \& {Travis}, J.~P. 1996, \aaps, 120, 537

\bibitem[{{Massaro} {et~al.}(2001){Massaro}, {Mantovani}, {Fanti}, {Nesci},
  {Tosti}, \& {Venturi}}]{2001A&A...374..435M}
{Massaro}, E., {Mantovani}, F., {Fanti}, R., {et~al.} 2001, \aap, 374, 435,
  \dodoi{10.1051/0004-6361:20010754}

\bibitem[{Montgomery(2012)}]{montgomery2012design}
Montgomery, D. 2012, Design and Analysis of Experiments, 8th Edition (John
  Wiley \& Sons, Incorporated).
\newblock \url{https://books.google.com.tr/books?id=XQAcAAAAQBAJ}

\bibitem[{{O'Neill} {et~al.}(2022){O'Neill}, {Kiehlmann}, {Readhead}, {Aller},
  {Blandford}, {Liodakis}, {Lister}, {Mr{\'o}z}, {O'Dea}, {Pearson}, {Ravi},
  {Vallisneri}, {Cleary}, {Graham}, {Grainge}, {Hodges}, {Hovatta},
  {L{\"a}hteenm{\"a}ki}, {Lamb}, {Lazio}, {Max-Moerbeck}, {Pavlidou}, {Prince},
  {Reeves}, {Tornikoski}, {Vergara de la Parra}, \&
  {Zensus}}]{2022ApJ...926L..35O}
{O'Neill}, S., {Kiehlmann}, S., {Readhead}, A.~C.~S., {et~al.} 2022, \apjl,
  926, L35, \dodoi{10.3847/2041-8213/ac504b}

\bibitem[{{Ostorero} {et~al.}(2006){Ostorero}, {Wagner}, {Gracia}, {Ferrero},
  {Krichbaum}, {Britzen}, {Witzel}, {Nilsson}, {Villata}, {Bach}, {Barnaby},
  {Bernhart}, {Carini}, {Chen}, {Chen}, {Ciprini}, {Crapanzano}, {Doroshenko},
  {Efimova}, {Emmanoulopoulos}, {Fuhrmann}, {Gabanyi}, {Giltinan},
  {Hagen-Thorn}, {Hauser}, {Heidt}, {Hojaev}, {Hovatta}, {Hroch}, {Ibrahimov},
  {Impellizzeri}, {Ivanidze}, {Kachel}, {Kraus}, {Kurtanidze},
  {L{\"a}hteenm{\"a}ki}, {Lanteri}, {Larionov}, {Lin}, {Lindfors}, {Munz},
  {Nikolashvili}, {Nucciarelli}, {O'Connor}, {Ohlert}, {Pasanen}, {Pullen},
  {Raiteri}, {Rector}, {Robb}, {Sigua}, {Sillanp{\"a}{\"a}}, {Sixtova},
  {Smith}, {Strub}, {Takahashi}, {Takalo}, {Tapken}, {Tartar}, {Tornikoski},
  {Tosti}, {Tr{\"o}ller}, {Walters}, {Wilking}, {Wills}, {Agudo}, {Aller},
  {Aller}, {Angelakis}, {Klare}, {K{\"o}rding}, {Strom}, {Ter{\"a}sranta},
  {Ungerechts}, \& {Vila-Vilar{\'o}}}]{2006A&A...451..797O}
{Ostorero}, L., {Wagner}, S.~J., {Gracia}, J., {et~al.} 2006, \aap, 451, 797,
  \dodoi{10.1051/0004-6361:20054075}

\bibitem[{{Otero-Santos} {et~al.}(2020){Otero-Santos}, {Acosta-Pulido},
  {Becerra Gonz{\'a}lez}, {Raiteri}, {Larionov}, {Pe{\~n}il}, {Smith},
  {Ballester Niebla}, {Borman}, {Carnerero}, {Castro Segura}, {Grishina},
  {Kopatskaya}, {Larionova}, {Morozova}, {Nikiforova}, {Savchenko},
  {Troitskaya}, {Troitsky}, {Vasilyev}, \& {Villata}}]{2020MNRAS.492.5524O}
{Otero-Santos}, J., {Acosta-Pulido}, J.~A., {Becerra Gonz{\'a}lez}, J.,
  {et~al.} 2020, \mnras, 492, 5524, \dodoi{10.1093/mnras/staa134}

\bibitem[{{Pandey} {et~al.}(2019){Pandey}, {Gupta}, {Wiita}, \&
  {Tiwari}}]{2019ApJ...871..192P}
{Pandey}, A., {Gupta}, A.~C., {Wiita}, P.~J., \& {Tiwari}, S.~N. 2019, \apj,
  871, 192, \dodoi{10.3847/1538-4357/aaf974}

\bibitem[{{Pichel} {et~al.}(2023){Pichel}, {Donzelli}, {Muriel}, {Rovero},
  {Rosa Gonz{\'a}lez}, {Vega}, {Aretxaga}, {Becerra Gonz{\'a}lez}, {Terlevich},
  {Terlevich}, \& {M{\'e}ndez-Abreu}}]{2023A&A...680A..52P}
{Pichel}, A., {Donzelli}, C., {Muriel}, H., {et~al.} 2023, \aap, 680, A52,
  \dodoi{10.1051/0004-6361/202245574}

\bibitem[{{Polednikova} {et~al.}(2016){Polednikova}, {Ederoclite}, {de Diego},
  {Cepa}, {Gonz{\'a}lez-Serrano}, {Bongiovanni}, {Oteo}, {Garc{\'\i}a},
  {P{\'e}rez-Mart{\'\i}nez}, {Pintos-Castro}, {Ram{\'o}n-P{\'e}rez}, \&
  {S{\'a}nchez-Portal}}]{2016MNRAS.460.3950P}
{Polednikova}, J., {Ederoclite}, A., {de Diego}, J.~A., {et~al.} 2016, \mnras,
  460, 3950, \dodoi{10.1093/mnras/stw1252}

\bibitem[{{Raiteri} {et~al.}(2003){Raiteri}, {Villata}, {Tosti}, {Nesci},
  {Massaro}, {Aller}, {Aller}, {Ter{\"a}sranta}, {Kurtanidze}, {Nikolashvili},
  {Ibrahimov}, {Papadakis}, {Krichbaum}, {Kraus}, {Witzel}, {Ungerechts},
  {Lisenfeld}, {Bach}, {Cim{\`o}}, {Ciprini}, {Fuhrmann}, {Kimeridze},
  {Lanteri}, {Maesano}, {Montagni}, {Nucciarelli}, \&
  {Ostorero}}]{2003A&A...402..151R}
{Raiteri}, C.~M., {Villata}, M., {Tosti}, G., {et~al.} 2003, \aap, 402, 151,
  \dodoi{10.1051/0004-6361:20030256}

\bibitem[{{Raiteri} {et~al.}(2017){Raiteri}, {Villata}, {Acosta-Pulido},
  {Agudo}, {Arkharov}, {Bachev}, {Baida}, {Ben{\'\i}tez}, {Borman}, {Boschin},
  {Bozhilov}, {Butuzova}, {Calcidese}, {Carnerero}, {Carosati}, {Casadio},
  {Castro-Segura}, {Chen}, {Damljanovic}, {D'Ammando}, {di Paola},
  {Echevarr{\'\i}a}, {Efimova}, {Ehgamberdiev}, {Espinosa}, {Fuentes},
  {Giunta}, {G{\'o}mez}, {Grishina}, {Gurwell}, {Hiriart}, {Jermak}, {Jordan},
  {Jorstad}, {Joshi}, {Kopatskaya}, {Kuratov}, {Kurtanidze}, {Kurtanidze},
  {L{\"a}hteenm{\"a}ki}, {Larionov}, {Larionova}, {Larionova}, {L{\'a}zaro},
  {Lin}, {Malmrose}, {Marscher}, {Matsumoto}, {McBreen}, {Michel}, {Mihov},
  {Minev}, {Mirzaqulov}, {Mokrushina}, {Molina}, {Moody}, {Morozova},
  {Nazarov}, {Nikolashvili}, {Ohlert}, {Okhmat}, {Ovcharov}, {Pinna},
  {Polakis}, {Protasio}, {Pursimo}, {Redondo-Lorenzo}, {Rizzi},
  {Rodriguez-Coira}, {Sadakane}, {Sadun}, {Samal}, {Savchenko}, {Semkov},
  {Skiff}, {Slavcheva-Mihova}, {Smith}, {Steele}, {Strigachev}, {Tammi},
  {Thum}, {Tornikoski}, {Troitskaya}, {Troitsky}, {Vasilyev}, \&
  {Vince}}]{2017Natur.552..374R}
{Raiteri}, C.~M., {Villata}, M., {Acosta-Pulido}, J.~A., {et~al.} 2017, \nat,
  552, 374, \dodoi{10.1038/nature24623}

\bibitem[{{Raiteri} {et~al.}(2021){Raiteri}, {Villata}, {Carosati},
  {Ben{\'\i}tez}, {Kurtanidze}, {Gupta}, {Mirzaqulov}, {D'Ammando}, {Larionov},
  {Pursimo}, {Acosta-Pulido}, {Baida}, {Balmaverde}, {Bonnoli}, {Borman},
  {Carnerero}, {Chen}, {Dhiman}, {Di Maggio}, {Ehgamberdiev}, {Hiriart},
  {Kimeridze}, {Kurtanidze}, {Lin}, {Lopez}, {Marchini}, {Matsumoto}, {Mujica},
  {Nakamura}, {Nikiforova}, {Nikolashvili}, {Okhmat}, {Otero-Santos}, {Rizzi},
  {Sakamoto}, {Semkov}, {Sigua}, {Stiaccini}, {Troitsky}, {Tsai}, {Vasilyev},
  \& {Zhovtan}}]{2021MNRAS.501.1100R}
{Raiteri}, C.~M., {Villata}, M., {Carosati}, D., {et~al.} 2021, \mnras, 501,
  1100, \dodoi{10.1093/mnras/staa3561}

\bibitem[{{Rani} {et~al.}(2010){Rani}, {Gupta}, {Strigachev}, {Bachev},
  {Wiita}, {Semkov}, {Ovcharov}, {Mihov}, {Boeva}, {Peneva}, {Spassov},
  {Tsvetkova}, {Stoyanov}, \& {Valcheva}}]{2010MNRAS.404.1992R}
{Rani}, B., {Gupta}, A.~C., {Strigachev}, A., {et~al.} 2010, \mnras, 404, 1992,
  \dodoi{10.1111/j.1365-2966.2010.16419.x}

\bibitem[{{Rani} {et~al.}(2013){Rani}, {Krichbaum}, {Fuhrmann}, {B{\"o}ttcher},
  {Lott}, {Aller}, {Aller}, {Angelakis}, {Bach}, {Bastieri}, {Falcone},
  {Fukazawa}, {Gabanyi}, {Gupta}, {Gurwell}, {Itoh}, {Kawabata}, {Krips},
  {L{\"a}hteenm{\"a}ki}, {Liu}, {Marchili}, {Max-Moerbeck}, {Nestoras},
  {Nieppola}, {Quintana-Lacaci}, {Readhead}, {Richards}, {Sasada}, {Sievers},
  {Sokolovsky}, {Stroh}, {Tammi}, {Tornikoski}, {Uemura}, {Ungerechts},
  {Urano}, \& {Zensus}}]{2013A&A...552A..11R}
{Rani}, B., {Krichbaum}, T.~P., {Fuhrmann}, L., {et~al.} 2013, \aap, 552, A11,
  \dodoi{10.1051/0004-6361/201321058}

\bibitem[{{Rani, B.} {et~al.}(2013){Rani, B.}, {Krichbaum, T. P.}, {Fuhrmann,
  L.}, {B\"ottcher, M.}, {Lott, B.}, {Aller, H. D.}, {Aller, M. F.},
  {Angelakis, E.}, {Bach, U.}, {Bastieri, D.}, {Falcone, A. D.}, {Fukazawa,
  Y.}, {Gabanyi, K. E.}, {Gupta, A. C.}, {Gurwell, M.}, {Itoh, R.}, {Kawabata,
  K. S.}, {Krips, M.}, {L\"ahteenm\"aki, A. A.}, {Liu, X.}, {Marchili, N.},
  {Max-Moerbeck, W.}, {Nestoras, I.}, {Nieppola, E.}, {Quintana-Lacaci, G.},
  {Readhead, A. C. S.}, {Richards, J. L.}, {Sasada, M.}, {Sievers, A.},
  {Sokolovsky, K.}, {Stroh, M.}, {Tammi, J.}, {Tornikoski, M.}, {Uemura, M.},
  {Ungerechts, H.}, {Urano, T.}, \& {Zensus, J. A.}}]{Rani2013}
{Rani, B.}, {Krichbaum, T. P.}, {Fuhrmann, L.}, {et~al.} 2013, A\&A, 552, A11,
  \dodoi{10.1051/0004-6361/201321058}

\bibitem[{{Romero} {et~al.}(2017){Romero}, {Boettcher}, {Markoff}, \&
  {Tavecchio}}]{2017SSRv..207....5R}
{Romero}, G.~E., {Boettcher}, M., {Markoff}, S., \& {Tavecchio}, F. 2017, \ssr,
  207, 5, \dodoi{10.1007/s11214-016-0328-2}

\bibitem[{{Romero} {et~al.}(1999){Romero}, {Cellone}, \&
  {Combi}}]{1999A&AS..135..477R}
{Romero}, G.~E., {Cellone}, S.~A., \& {Combi}, J.~A. 1999, \aaps, 135, 477,
  \dodoi{10.1051/aas:1999184}

\bibitem[{{Scargle}(1982)}]{1982ApJ...263..835S}
{Scargle}, J.~D. 1982, \apj, 263, 835, \dodoi{10.1086/160554}

\bibitem[{{Schlafly} \& {Finkbeiner}(2011)}]{2011ApJ...737..103S}
{Schlafly}, E.~F., \& {Finkbeiner}, D.~P. 2011, \apj, 737, 103,
  \dodoi{10.1088/0004-637X/737/2/103}

\bibitem[{Schmidt {et~al.}(2011)Schmidt, Rix, Shields, Knecht, Hogg, Maoz, \&
  Bovy}]{Schmidt_2012}
Schmidt, K.~B., Rix, H.-W., Shields, J.~C., {et~al.} 2011, The Astrophysical
  Journal, 744, 147, \dodoi{10.1088/0004-637X/744/2/147}

\bibitem[{{Simonetti} {et~al.}(1985){Simonetti}, {Cordes}, \&
  {Heeschen}}]{1985ApJ...296...46S}
{Simonetti}, J.~H., {Cordes}, J.~M., \& {Heeschen}, D.~S. 1985, \apj, 296, 46,
  \dodoi{10.1086/163418}

\bibitem[{{Stalin} {et~al.}(2006){Stalin}, {Gopal-Krishna}, {Sagar}, {Wiita},
  {Mohan}, \& {Pandey}}]{2006MNRAS.366.1337S}
{Stalin}, C.~S., {Gopal-Krishna}, {Sagar}, R., {et~al.} 2006, \mnras, 366,
  1337, \dodoi{10.1111/j.1365-2966.2005.09939.x}

\bibitem[{{Tripathi} {et~al.}(2023){Tripathi}, {Gupta}, {Takey}, {Bachev},
  {Vince}, {Strigachev}, {Kushwaha}, {Elhosseiny}, {Wiita}, {Damljanovic},
  {Dhiman}, {Fouad}, {Gaur}, {Gu}, {Hamed}, {Kishore}, {Kurtenkov}, {Rastogi},
  {Semkov}, {Zead}, \& {Zhang}}]{2023MNRAS.tmp.3427T}
{Tripathi}, T., {Gupta}, A.~C., {Takey}, A., {et~al.} 2023, \mnras,
  \dodoi{10.1093/mnras/stad3574}

\bibitem[{{Urry} \& {Padovani}(1995)}]{1995PASP..107..803U}
{Urry}, C.~M., \& {Padovani}, P. 1995, \pasp, 107, 803, \dodoi{10.1086/133630}

\bibitem[{{Villata} {et~al.}(1998){Villata}, {Raiteri}, {Lanteri}, {Sobrito},
  \& {Cavallone}}]{1998A&AS..130..305V}
{Villata}, M., {Raiteri}, C.~M., {Lanteri}, L., {Sobrito}, G., \& {Cavallone},
  M. 1998, \aaps, 130, 305, \dodoi{10.1051/aas:1998415}

\bibitem[{{Villata} {et~al.}(2002){Villata}, {Raiteri}, {Kurtanidze},
  {Nikolashvili}, {Ibrahimov}, {Papadakis}, {Tsinganos}, {Sadakane}, {Okada},
  {Takalo}, {Sillanp{\"a}{\"a}}, {Tosti}, {Ciprini}, {Frasca}, {Marilli},
  {Robb}, {Noble}, {Jorstad}, {Hagen-Thorn}, {Larionov}, {Nesci}, {Maesano},
  {Schwartz}, {Basler}, {Gorham}, {Iwamatsu}, {Kato}, {Pullen}, {Ben{\'\i}tez},
  {de Diego}, {Moilanen}, {Oksanen}, {Rodriguez}, {Sadun}, {Kelly}, {Carini},
  {Miller}, {Catalano}, {Dultzin-Hacyan}, {Fan}, {Ishioka}, {Karttunen},
  {Kein{\"a}nen}, {Kudryavtseva}, {Lainela}, {Lanteri}, {Larionova},
  {Matsumoto}, {Mattox}, {Montagni}, {Nucciarelli}, {Ostorero},
  {Papamastorakis}, {Pasanen}, {Sobrito}, \& {Uemura}}]{2002A&A...390..407V}
{Villata}, M., {Raiteri}, C.~M., {Kurtanidze}, O.~M., {et~al.} 2002, \aap, 390,
  407, \dodoi{10.1051/0004-6361:20020662}

\bibitem[{{Villata} {et~al.}(2008){Villata}, {Raiteri}, {Larionov},
  {Kurtanidze}, {Nilsson}, {Aller}, {Tornikoski}, {Volvach}, {Aller},
  {Arkharov}, {Bach}, {Beltrame}, {Bhatta}, {Buemi}, {B{\"o}ttcher},
  {Calcidese}, {Carosati}, {Castro-Tirado}, {da Rio}, {di Paola}, {Dolci},
  {Forn{\'e}}, {Frasca}, {Hagen-Thorn}, {Heidt}, {Hiriart}, {Jel{\'\i}nek},
  {Kimeridze}, {Konstantinova}, {Kopatskaya}, {Lanteri}, {Leto}, {Ligustri},
  {Lindfors}, {L{\"a}hteenm{\"a}ki}, {Marilli}, {Nieppola}, {Nikolashvili},
  {Pasanen}, {Ragozzine}, {Ros}, {Sigua}, {Smart}, {Sorcia}, {Takalo},
  {Tavani}, {Trigilio}, {Turchetti}, {Uckert}, {Umana}, {Vercellone}, \&
  {Webb}}]{2008A&A...481L..79V}
{Villata}, M., {Raiteri}, C.~M., {Larionov}, V.~M., {et~al.} 2008, \aap, 481,
  L79, \dodoi{10.1051/0004-6361:200809552}

\bibitem[{{Volvach} {et~al.}(2012){Volvach}, {Volvach}, {Bychkova},
  {Kardashev}, {Larionov}, {Vlasjuk}, {Spiridonova}, {Lachteenmaki},
  {Tornikoski}, {Nieppola}, {Aller}, \& {Aller}}]{2012ARep...56..275V}
{Volvach}, A.~E., {Volvach}, L.~N., {Bychkova}, V.~S., {et~al.} 2012, Astronomy
  Reports, 56, 275, \dodoi{10.1134/S1063772912030079}

\bibitem[{{Wagner} \& {Witzel}(1995)}]{1995ARA&A..33..163W}
{Wagner}, S.~J., \& {Witzel}, A. 1995, \araa, 33, 163,
  \dodoi{10.1146/annurev.aa.33.090195.001115}

\bibitem[{{Wagner} {et~al.}(1996){Wagner}, {Witzel}, {Heidt}, {Krichbaum},
  {Qian}, {Quirrenbach}, {Wegner}, {Aller}, {Aller}, {Anton}, {Appenzeller},
  {Eckart}, {Kraus}, {Naundorf}, {Kneer}, {Steffen}, \&
  {Zensus}}]{1996AJ....111.2187W}
{Wagner}, S.~J., {Witzel}, A., {Heidt}, J., {et~al.} 1996, \aj, 111, 2187,
  \dodoi{10.1086/117954}

\bibitem[{{Wiita}(1996)}]{1996ASPC..110...42W}
{Wiita}, P.~J. 1996, in Astronomical Society of the Pacific Conference Series,
  Vol. 110, Blazar Continuum Variability, ed. H.~R. {Miller}, J.~R. {Webb}, \&
  J.~C. {Noble}, 42

\bibitem[{{Yuan} {et~al.}(2017){Yuan}, {Fan}, {Tao}, {Qian}, {Costantin},
  {Xiao}, {Pei}, \& {Lin}}]{2017A&A...605A..43Y}
{Yuan}, Y.-H., {Fan}, J.-h., {Tao}, J., {et~al.} 2017, \aap, 605, A43,
  \dodoi{10.1051/0004-6361/201630338}

\bibitem[{{Zhang} {et~al.}(2022){Zhang}, {Li}, {Giannios}, {Guo}, {Thiersen},
  {B{\"o}ttcher}, {Lewis}, \& {Venters}}]{2022ApJ...924...90Z}
{Zhang}, H., {Li}, X., {Giannios}, D., {et~al.} 2022, \apj, 924, 90,
  \dodoi{10.3847/1538-4357/ac3669}

\end{thebibliography}
\bibliographystyle{aasjournal}

\end{document}